\documentstyle{article}

\def\bea{\begin{eqnarray}}
\def\eea{\end{eqnarray}}
\def\nn{\nonumber}

\def\beq{\begin{equation}}
\def\eeq{\end{equation}}
\def\ba{\beq\new\begin{array}{c}}
\def\ea{\end{array}\eeq}
\def\be{\ba}
\def\ee{\ea}
\def\stackreb#1#2{\mathrel{\mathop{#2}\limits_{#1}}}
\def\Tr{{\rm Tr}}
\def\res{{\rm res}}

\def\rank{{\rm rank}}

\def\2{{1\over 2}}

\def\Bf#1{\mbox{\boldmath $#1$}}
\def\balpha{{\Bf\alpha}}

\def\bsigma{{\bfit\sigma}}

\begingroup\ifx\undefined\newsymbol \else\def\input#1 {\endgroup}\fi
\input amssym.def \relax
\input amssym
\newfont{\hr}{msbm10}
\newfont{\ams}{msam10}
\font\teneufm=cmmib10
\font\seveneufm=cmmib7
\font\fiveeufm=cmmib5
\def\bfit#1{{\textfont1=\teneufm\scriptfont1=\seveneufm
\scriptscriptfont1=\fiveeufm
\mathchoice{\hbox{$\displaystyle#1$}}{\hbox{$\textstyle#1$}}
{\hbox{$\scriptstyle#1$}}{\hbox{$\scriptscriptstyle#1$}}}}

\makeatletter
\newdimen\normalarrayskip              
\newdimen\minarrayskip                 
\normalarrayskip\baselineskip
\minarrayskip\jot
\newif\ifold             \oldtrue            \def\new{\oldfalse}
\def\arraymode{\ifold\relax\else\displaystyle\fi} 
\def\eqnumphantom{\phantom{(\theequation)}}     
\def\@arrayskip{\ifold\baselineskip\z@\lineskip\z@
     \else
     \baselineskip\minarrayskip\lineskip2\minarrayskip\fi}
\def\@arrayclassz{\ifcase \@lastchclass \@acolampacol \or
\@ampacol \or \or \or \@addamp \or
   \@acolampacol \or \@firstampfalse \@acol \fi
\edef\@preamble{\@preamble
  \ifcase \@chnum
     \hfil$\relax\arraymode\@sharp$\hfil
     \or $\relax\arraymode\@sharp$\hfil
     \or \hfil$\relax\arraymode\@sharp$\fi}}
\def\@array[#1]#2{\setbox\@arstrutbox=\hbox{\vrule
     height\arraystretch \ht\strutbox
     depth\arraystretch \dp\strutbox
     width\z@}\@mkpream{#2}\edef\@preamble{\halign
\noexpand\@halignto
\bgroup \tabskip\z@ \@arstrut \@preamble \tabskip\z@ \cr}%
\let\@startpbox\@@startpbox \let\@endpbox\@@endpbox
  \if #1t\vtop \else \if#1b\vbox \else \vcenter \fi\fi
  \bgroup \let\par\relax
  \let\@sharp##\let\protect\relax
  \@arrayskip\@preamble}
\def\eqnarray{\stepcounter{equation}%
              \let\@currentlabel=\theequation
              \global\@eqnswtrue
              \global\@eqcnt\z@
              \tabskip\@centering
              \let\\=\@eqncr
              $$%
 \halign to \displaywidth\bgroup
    \eqnumphantom\@eqnsel\hskip\@centering
    $\displaystyle \tabskip\z@ {##}$%
    \global\@eqcnt\@ne \hskip 2\arraycolsep
         $\displaystyle\arraymode{##}$\hfil
    \global\@eqcnt\tw@ \hskip 2\arraycolsep
         $\displaystyle\tabskip\z@{##}$\hfil
         \tabskip\@centering
    &{##}\tabskip\z@\cr}

\begin{document}

\begin{titlepage}
\setcounter{footnote}0
\begin{center}
\hfill FIAN/TD-16/96\\
\hfill ITEP/TH-47/96\\
\vspace{0.3in}
{\LARGE\bf Non-perturbative Quantum Theories and Integrable Equations}
\\
\bigskip
\bigskip
\bigskip
{\Large A.Marshakov
\footnote{E-mail address: mars@lpi.ac.ru}}
\\
\bigskip
{\it Theory Department,  P. N. Lebedev Physics
Institute , Leninsky prospect, 53, Moscow,~117924, Russia\\
and ITEP, Moscow 117259, Russia}
\end{center}
\bigskip \bigskip

\begin{abstract}
I review the appearance of classical integrable systems as an effective
tool for the description of non-perturbative exact results in quantum
string and gauge theories. Various aspects of this relation: spectral
curves, action-angle variables, Whitham deformations and associativity
equations are considered separately demonstrating hidden parallels
between topological $2d$ string theories and naively non-topological $4d$
theories. The proofs are supplemented by explicit illustrative examples.
\end{abstract}

\end{titlepage}

\newpage
\setcounter{footnote}0

\section{Introduction}

In this paper I will try to describe recent results in modern
quantum field theory and string theory related to the exact construction of
{\it non-\-perturbative} solutions and their formulation in terms of
integrable systems.
Starting from quantum field theory the problem
is to compute exactly the quantum spectrum and the correlation functions
\beq\label{corr}
\langle \Phi _{k_1}\dots\Phi _{k_n}\rangle = \int D\varphi e^{-{1\over\hbar}
\int {\cal L}(\varphi,\partial _{\mu}\varphi )}
\Phi _{{\bf k}_1}(\varphi ,\partial _{\mu}\varphi ,\dots )\dots\Phi_{{\bf k}_n}
(\varphi ,
\partial _{\mu}\varphi ,\dots ) \equiv F_{{\bf k}_1,\dots ,{\bf k}_n}(t,u)
\eeq
as functions of coupling constants $\{t_k\}$, quantum characteristics of
physical operators $\{{\bf k}_1,\dots ,{\bf k}_n\}$ and possible parameters
of classical solutions $\{ u_{\alpha}\} $ or {\it moduli} of the theory.

Unfortunately almost never the integral in (\ref{corr}) can be computed
exactly. Up to now the only positive experience of computing (\ref{corr})
comes from perturbative calculations and/or lattice regularization of
(\ref{corr})
resulting usually in some approximations being quite far from the exact answer.
In string theory the situation is even more complicated since there is no real
target-\-space formulation of the problem: the Polyakov path integral
\be\label{pol}
{\cal F}(\Lambda) = \sum_{genus} \Lambda ^g F_g
\ \ \ \ \ \ \
F_g = \int _{\Sigma_g} Dg Dx e^{-{1\over \hbar} \int {\cal L}(x,g)}
\ee
is
 by definition only a perturbative formulation.

Recent investigations showed however (see for example
\cite{ds,FKN1,Kontsevich,GKM,LGGKM,SW1,SW2,VaWi,GKMMM}) that sometimes
there exists a way to find explicitly the exact nonperturbative results
(spectrum, correlation functions, effective actions) even in the quantum
theories which can not be considered as {\it quantum integrable} models
(see for example \cite{FT,qi} and references therein)
at least in conventional and naive sense. In contrast to "naive"
quantum integrable models where usually it is (infinite-dimensional) algebra
of quantum symmetry which allows one to calculate the spectrum
and the correlation functions, there is no even to such extent
"straightforward"
way in the theories I am going to discuss. The intention to study these
particular models is caused by hope that they are not so far away from
the realistic quantum gauge and string theories where the solution to the
basic problems of confinement and quantum gravity is looked for, being
on the other hand {\it solvable} at least in the sense to be discussed below.

The starting point is that even in this class of models
there are still no {\it direct} ways of solution -- like there are no such
ways for in many other models of quantum field and string theory.
Amazingly enough, the indirect arguments show that all existing examples
of the exact non-\-perturbative solutions have almost identically the
{\it same effective} formulation which can be presented in terms of effective
completely integrable model forgetting about many properties of the bare theory
\footnote{In particular there
is no real distinction between the theories living in different space-time
dimensions.}. The low-energy limit of such models is also described by a sort
of {\it topological} theory -- which is not topological in naive sense still
having lots of properties of conventional (essentially $2d$)
topological models.

It is still not possible to describe the mechanism which obtains the
final results starting directly from the Lagrangian formulation
\footnote{One hope is that the arising equations of hydrodynamical type
might be considered as generalizations of the renormalization-\-group
technique of the
perturbation theory. Indeed it is known that in the perturbation theory the
scale dependence of quantum correlation functions is related with their
dependence on the coupling constants by means of the first order equation
$$
\left( {d\over d\log\mu} - \sum \beta _i(g){\partial\over\partial g_i}\right)
F(g;\mu ) = 0
$$
Naively in the exact solution there is no scale dependence, so the
resulting equation could be of the form $\sum \beta _i(g){\partial F\over
\partial g_i} = 0$, or the derivatives over $\mu $ could be replaced by the
derivatives over moduli $\{ u_{\alpha }\}$. Even accepting this it is necessary
to stress that there is no known way to define $\beta $-function beyond the
perturbation theory (i.e. not using the scale dependence $\beta _i(g) =
{\partial g_i\over \partial\log\mu}$). However, one should also remember that
some of
the integrable equations -- for example the associativity equations discussed
in the last section look like having a purely {\it string} theory nature.},
however in what follows I will try to believe that it exists and formulate the
basic features of the effective description, which attract a lot of interest
from the point of view of integrable systems themselves.

First, the parameters of the theory -- or its moduli come -- usually from the
(gauge-invariant) low-energy values of the (background) fields in the
target-space.
For example, in $4d$ supersymmetric gauge theories these are in part given by
the v.e.v.'s of the Higgs fields $h_k = {1\over k}\langle\Tr\Phi ^k\rangle$,
in string theories -- by
the moduli of the target-\-space metric (e.g. K\"ahler or
complex structures), gauge fields (e.g. moduli of flat connections or
selfdual gauge fields) etc. The problem
itself is to find the (exact nonperturbative) dependence of the physical
objects upon the moduli parameters
\footnote{Of course, in addition to the parametric dependence of moduli
themselves the physical quantities can depend on the topological (discrete)
characteristics of moduli spaces, moreover in the simplest topological string
models only this dependence is essential and the correlation functions can be
just {\it numbers}.}.
The idea that the moduli of (some) target-\-space
theories can be identified with the moduli of the complex manifolds (in
particular having the property that their
different values can be treated as equivalent if related by an action
of some (discrete) group) is usually called as duality concept
(see for example \cite{dualn})while the action of group itself --
a {\it duality} map.

A crucial point is that the class of the theories we discuss
is distinguished by the {\it holomorphic} dependence of the (complex)
moduli parameters -- i.e. there exists a complex structure on the moduli space
and only functions with "good" global behaviour enter the game. This goes back to
the holomorphic structures arising in the instantonic calculus and to the
Belavin-Knizhnik theorem in string theory leading to drastical
simplifications removing the possible ambiguity in the form of the exact
answer. In the known examples the moduli of the theory appear to be (a subspace
in) the moduli space of complex structures of the {\it target-\-space} spectral
curves $\Sigma $
\footnote{From the general properties of string theories one can expect that
the same picture exists when the moduli of the theory are effectively described
as moduli of higher dimensional {\it complex} manifolds ($K3$, the Calabi-\-Yau
3-folds etc.). The particular case considered in this text corresponds to a
specific situation when the CY manifold effectively degenerates into
$1_{\bf C}d$ complex curve $\Sigma $ \cite{VafaC}.
}. It turns out, that {\it nonperturbatively} the target-\-space
spectral curve acquires a nontrivial topological structure (being just defined
locally -- as a
sphere of the scale parameter in perturbation theory) and complex
structure on the spectral curve is parameterized by moduli of the
theory. This additional complicated structure
means that "stringy" nature plays an important role in the nonperturbative
formulation and strings (and $D$-branes) wrapping along topologically
nontrivial directions produce important effects in the exact effective
formulation \cite{VafaC} while
perturbatively the spectral curve can be tested only "locally".

To be precise, the relation between the non-\-perturbative solutions
and integrable theories was described in detail for some topological
$2d$ theories and so far for the Seiberg-Witten (SW) solutions only for
particular families of
models: the most known is the ${\cal N}=2$ SYM theory with one ($N_a=1$)
``matter" ${\cal N}=2$ hypermultiplet in the adjoint representation of
the gauge group $G$ -- which is known to be related to the
Calogero-Moser family of integrable systems \cite{WiDo,Mart,GM,MartWa,IM}.
When the hypermultiplet decouples (its mass becomes infinite), the dimensional
transmutation takes place and the pure gauge $4d$ ${\cal N}=2$ SYM theory gets
associated with the Toda-chain model \cite{GKMMM}.
It is also known \cite{M4,AN} that the
$N_c = 3$, $N_f = 2$ curve can be associated with the
Goryachev-\-Chaplygin top while a natural suggestion for the whole
${\cal N}=2$ SQCD family is to associate it with the
well-known family of integrable systems -- (in general inhomogeneous) $sl(2)$
spin chains, with the Toda chain (pure gauge model) being again a limiting
case. The crucial motivation for such a suggestion \cite{M4}
is the specific quadratic form of the spectral equations, derived in
\cite{SW2,fumat}
\footnote{Another possibility \cite{KriPho} is to stay within the frames
of the Toda chain modifying only the boundary conditions (i.e. changing $T$
operator for the same ${\cal L}$). This however seems to be not enough to
consider the situation with $N_f > N_c$.}.

Now, once the solutions are formulated in terms of periods of some
differentials on a complex spectral curve an integrable system
(moreover the particular class of systems of KP/Toda type -- where the
Liouville torus is restricted to be a real section of Jacobian of a
complex curve) arises more or less by definition due to the Krichever
construction (see \cite{DKN} and references therein). This {\it very useful}
observation leads to a possibility of applying rather simple technique of Lax
pairs, spectral
curves, symplectic forms and $\tau $-functions to naively infinite-dimensional
quantum field and string theories. Again it means that the effective
{\it non-\-perturbative} theory can be formulated in the language of
(finite-\-dimensional) integrable systems, i.e. only small part of all
degrees of freedom is essential for the exact non-\-perturbative behaviour.

The plan of this paper looks as follows. First, I consider the {\it fact} of
appearance of curves and integrable systems and formulate the string/QFT
results in these terms. Next, in sect.3 it is shown that a completely
integrable system arises after one introduces \cite{DKN} a {\it meromorphic}
1-form $dS$ whose derivatives
\be\label{hol}
{\partial dS\over\partial h_k}\cong d\omega _k
\ee
give {\it holomorphic} differentials and it is explained that the generating
1-form
defines a completely integrable system on a symplectic manifold endowed with
symplectic form $\Omega \sim \delta dS $ which in all cases we discuss can be
essentially written as
\be\label{dS}
dS = \lambda d\log w = \Tr{\cal L}d\log T
\ \ \ \ \ \ \ \
\Omega = \delta\lambda\wedge\delta\log w = \Tr\delta{\cal L}\wedge\delta\log T
\ee
The symplectic form (\ref{dS}) is defined by the eigenvalues of {\it two}
operators. Quasiclassically their common spectrum defines the spectral curve.
The symplectomorphisms of (\ref{dS}) can be
considered as transformations between the {\it dual} integrable systems
with the generating function $S = \sum _k\int ^{\gamma _k}dS$.

Indeed, writing (\ref{dS}) more explicitly one finds that $\Omega = \sum _k
\left.\delta\lambda\wedge\delta\log w\right|_{\gamma _k}=\sum _k\delta h_k
\wedge\delta\phi _k$ and the Hamiltonian flows provided by $h_k$ produce a
completely integrable system on Jacobian with co-\-ordinates $\{\phi _k\}$.
Next, in sect.4 another property of the same 1-differential $dS$ is
discussed -- $dS$ (\ref{dS}) plays the role of the generating differential of
the Whitham
hierarchy describing the flows in moduli space around a point, corresponding to
a finite-gap solution. The basic features of the Whitham hierarchy are
presented and some explicit examples of its solutions are considered.
Finally in sect.5 the central object containing the most complete information
about the integrable system $\log{\cal T} = \log{\cal T}_0
+ \log{\cal T}_{\theta}$ is discussed. Its part (a logariphm of a
quasiclassical $\tau $- function) restricted to the dependence
on moduli
\be\label{prepot}
\left.\log{\cal T}_0\right|_{moduli} \equiv {\cal F}
\ee
(usually called a {\it prepotential}) is defined in terms of
the "periods" $t_k = \oint _{C_k}dS$ or $t_{\alpha}
= \res _{P_{\alpha}}(\lambda ^{-\alpha}dS)$ and intersection form
$T_{ik} = \int _{\Sigma}d\omega _i\wedge d\omega _j$ on $\Sigma$.
The main statement of sect.5 is that the prepotential
${\cal F} = \log{\cal T}_0$ in general satisfies the
associativity equation having the form (for the matrices ${\cal F}_{ijk} =
{\partial ^3{\cal F}\over
\partial t_i\partial t_j\partial t_k} \equiv ({\cal F}_i)_{jk}$)
\be\label{assoc}
{\cal F}_i{\cal F}_j^{-1}{\cal F}_k = {\cal F}_k{\cal F}_j^{-1}{\cal F}_i
\ \ \ \ \ \forall i,j,k
\ee
In Conclusion some open questions are listed and the relation with string
dualities is discussed.

\section{Curves and Integrable systems}

The simplest example of the nonperturbative solutions is when they exist
in {\it explicit} form, being related to the rational spectral curve
$\Sigma $ where only the times connected to the residues are valid,
the prepotential ${\cal F} = {t_1^3\over 6} + \dots$ is polynomial in times
and the eigenvalues of two operators in (\ref{dS}) are polynomial functions on
Riemann sphere $CP^1$. The topological correlators are
{\it numbers} and count the intersection indices on moduli space -- this
is an example of {\it topological} gravity \cite{WDVV,Kontsevich,GKM}.

The case of {\it physical} ($c < 1$ or $pq$-) gravity is known much less
explicitly and correspond already to nontrivial spectral curves
$\Sigma _{g={(p-1)(q-1)\over 2}}$ \cite{KNMS}
\footnote{This is however the case where the exact
form of the {\it duality} transformation -- relating the partition functions
in the dual points -- is known exactly, having the form of a Fourier transform
with the exponent $S = \int ^{\lambda}dS$.}. Formally the topological $2d$
theories with the "target-\-space" higher genus spectral curves were
constructed in \cite{KriW,Dubtop}.

The higher genus complex curves arise also in $4d$ SUSY gauge theories --
the field theory limit $\alpha '\rightarrow 0$ of $c > 1$ supersymmetric
string theories
\beq\label{ym}
{\cal L} =\int d^4\theta F(\Phi_i) = \dots
 {1\over g^2}\Tr F_{\mu\nu}^2 + i\theta \Tr F_{\mu\nu}
{\tilde F}_{\mu\nu} + \dots
\eeq
(the superfield $\Phi_i = \varphi^i + \theta\sigma_{\mu\nu}
\tilde\theta G_{\mu\nu}^i + \ldots$) where the nonperturbative exact solution
is formally defined as a map
\be\label{ymval}
G,\tau ,h_k \rightarrow a_i,\ a_i^D,\ a_i^D =
{\partial{\cal F}\over\partial a_i}
\ \ \ \ \ \ \
T_{ij} = {\partial ^2{\cal F}\over \partial a_i\partial a_j} \ \ \ \ \ \
M \sim |{\bf na} + {\bf ma}^D|
\ee
$G$ is gauge group, $\tau$ -- the UV coupling
constant, $h_k={1\over k}\langle\Tr\Phi ^k\rangle$ -- the v.e.v.'s
of the Higgs field)
and an elegant description in terms
of $\Sigma _{g = \rank G}$ with $h_k$ parameterizing some of the
(in most cases hyperelliptic) moduli
of complex structures. The periods of meromorphic 1-form (\ref{dS})
$a_i = \oint_{A_i} dS$, $a_i^D = \oint_{B_i} dS$ determine the BPS massive
spectrum, $a_i^D=\frac{\partial{\cal F}}{\partial a^i}$ the prepotential
${\cal F}$ (giving the low-energy effective action) and, thus, the {\it set}
of low-energy coupling constants $T_{ij} = \frac{\partial^2{\cal F}}
{\partial a_i\partial a_j} =
\frac{\partial a^D_i}{\partial a_j}$.
The curves $\Sigma _{g = \rank G}$ are special spectral curves
\footnote{They correspond to the $g$-parametric families of complex curves
in the moduli space of total dimension $3g-3$.} of the
nontrivial finite-gap solutions to the periodic Toda-chain problem and its
natural deformations.

\subsection{Toda chain: $N_c\times N_c$ versus $2\times 2$
representation}

We start our analysis from the simplest Toda-chain model, which in the
framework of the nonperturbative solutions
corresponds to the $4d$ {\it pure} gauge ${\cal N}=2$ supersymmetric Yang-Mills
theory \cite{sun,GKMMM}. The periodic problem in this model can be formulated
in two different ways, which will be further deformed into two different
directions. These deformations
are hypothetically related to the two different couplings of the $4d$
theory by adding the adjoint and fundamental matter
${\cal N}=2$ hypermultiplets correspondingly.

The Toda chain system can be defined by the equations of motion
\be\label{Todaeq}
\frac{\partial q_i}{\partial t} = p_i \ \ \ \ \
\frac{\partial p_i}{\partial t} = e^{q_{i+1} -q_i}- e^{q_i-q_{i-1}}
\ee
where one assumes (for the periodic problem with the ``period" $N_c$) that
$q_{i+N_c} = q_i$ and
$p_{i+N_c} = p_i$. It is an integrable system, with $N_c$
Poisson-commuting Hamiltonians, $h_1^{TC} = \sum p_i$, $h_2^{TC} =
\sum\left(\frac{1}{2}p_i^2 + e^{q_i-q_{i-1}}\right)$, etc. As
any finite-gap solution the periodic problem in Toda chain is described
in terms of (the eigenvalues and the eigenfunctions of) two operators:
the Lax operator ${\cal L}$ (or the auxiliary linear problem for (\ref{Todaeq}))
\be\label{laxtoda}
\lambda\psi ^{\pm}_n =
\sum _k {\cal L}_{nk}\psi ^{\pm}_k =
e^{{1\over 2}(q_{n+1}-q_n)}\psi ^{\pm}_{n+1} + p_n\psi ^{\pm}_n +
e^{{1\over 2}(q_n-q_{n-1})}
\psi ^{\pm}_{n-1}
\ (=  \pm {\partial\over\partial t}\psi ^{\pm}_n)
\ee
and the second is a {\it monodromy} or shift operator in a discrete
variable -- the number of a particle
\be\label{T-op}
Tq_n = q_{n+N_c}\ \ \ \ \ \ Tp_n = p_{n+N_c}\ \ \ \ \ \ \ \
T\psi_n = \psi_{n+N_c}
\ee
The common spectrum of these two operators
\be\label{spec}
{\cal L}\psi = \lambda\psi \ \ \ \ \ T\psi = w\psi\ \ \ \ \ \ [{\cal L},T]=0
\ee
means that there exists a relation between them ${\cal P}({\cal L},T) = 0$
which can be strictly formulated in terms of spectral curve $\Sigma$:
${\cal P}(\lambda,w) = 0$. The
generation function for these Hamiltonians can be written in terms of
${\cal L}$ and $T$ operators and the Toda chain possesses two
essentially different formulations of this kind.

In the first version (which can be considered as a limiting case of
Hitchin system \cite{Hi}), the Lax operator (\ref{laxtoda}) is written in the
basis of the $T$-operator eigenfunctions and becomes the $N_c\times N_c$
matrix,
\be\label{LaxTC}
{\cal L}^{TC}(w) =
\left(\begin{array}{ccccc}
 p_1 & e^{{1\over 2}(q_1-q_2)} & 0 & & we^{{1\over 2}(q_1-q_{N_c})}\\
e^{{1\over 2}(q_2-q_1)} & p_2 & e^{{1\over 2}(q_2 - q_3)} & \ldots & 0\\
0 & e^{{1\over 2}(q_3-q_2)} & p_3 & & 0 \\
 & & \ldots & & \\
\frac{1}{w}e^{{1\over 2}(q_{N_c}-q_1)} & 0 & 0 & & p_{N_c}
\end{array} \right)
\ee
defined on the two-punctured sphere. The Poisson brackets
$\{p_i,q_j\} = \delta_{ij}$ imply that it satisfies a PB relation
\be
\left\{ {\cal L}^{TC}(w)\stackrel{\otimes}{,}
{\cal L}^{TC}(w')\right\} =
\left[ {\cal R}(w,w'), {\cal L}^{TC}(w)\otimes {\bf 1} + {\bf 1}\otimes
{\cal L}^{TC}(w')\right]
\ee
with the {\it numeric} trigonometric
${\cal R}$-matrix,
\be
{\cal R}(w,w')={w\sum
\left(
\delta_{i,i+1}\otimes \delta_{i+1,i} \right)
+\left( w'\sum \delta_{i+1,i}\otimes \delta_{i,i+1}   \right)
\over w-w'}
\ee
and the eigenvalues of the Lax operator defined from the spectral
equation
\be\label{SpeC}
{\cal P}(\lambda,w) = \det_{N_c\times N_c}\left({\cal L}^{TC}(w) -
\lambda\right) = 0
\ee
are Poisson-commuting with each other.
Substituting the explicit expression (\ref{LaxTC}) into (\ref{SpeC}),
one gets \cite{KriDu}:
\be\label{fsc-Toda}
w + \frac{1}{w} = 2P_{N_c}(\lambda )
\ee
or
\be\label{hypelTC}
y^2 = P_{N_c}^2(\lambda ) - 1 \ \ \ \ \ \ \ 2y = w - {1\over w}
\ee
where $P_{N_c}(\lambda )$ is a polynomial of degree $N_c$,
with the coefficients being the Schur polynomials of the Hamiltonians
$h_k = \sum_{i=1}^{N_c} p_i^k + \ldots$:
\be
P_{N_c}(\lambda ) = \left( \lambda ^{N_c} + h_1 \lambda ^{N_c-1} +
\frac{1}{2}(h_2-h_1^2)\lambda ^{N_c-2} + \ldots \right)
\ee
The spectral equation depends only on the mutually Poisson-commuting
combinations of the dynamical variables --
the Hamiltonians or better action variables --
parameterizing (a subspace in the) moduli space of the complex structures of
the hyperelliptic curves $\Sigma^{TC}$ of genus $N_c - 1 = \rank SU(N_c)$.

An alternative description of the same system arises when one {\it solves}
explicitly the auxiliary linear problem (\ref{laxtoda}) which is a
{\it second}-order
difference equation -- to solve which one just rewrites it as
${\tilde\psi}_{i+1}
= L^{TC}_i(\lambda ){\tilde\psi}_i$ (after a simple "gauge"
transformation) with the help of a chain of $2\times 2$ Lax matrices \cite{FT}
\be\label{LTC}
L^{TC}_i(\lambda) =
\left(\begin{array}{cc} p_i + \lambda & e^{q_i} \\ e^{-q_i} & 0
\end{array}\right), \ \ \ \ \ i = 1,\dots ,N_c
\ee
These matrices obey the {\it
quadratic} $r$-matrix Poisson relations \cite{Skl}
\be\label{quadrP} \left\{
L^{TC}_i(\lambda)\stackrel{\otimes}{,}L_j^{TC}(\lambda')\right\} =
\delta_{ij} \left[ r(\lambda - \lambda'), L^{TC}_i(\lambda)\otimes
L^{TC}_j(\lambda')\right]
\ee
with the ($i$-independent!) numerical rational
$r$-matrix satisfying the classical Yang-Baxter
equation $r(\lambda) =
\frac{1}{\lambda } \sum_{a=1}^3 \sigma_a\otimes \sigma^a$.
As a consequence, the transfer matrix (generally defined for the
inhomogeneous lattice with inhomogenities $\lambda_i$'s)
\be\label{monomat}
T_{N_c}(\lambda) =
\prod_{1\ge i\ge N_c}^{\curvearrowleft} L_i(\lambda - \lambda_i)
\ee
satisfies the same Poisson-\-bracket relation
\be
\left\{ T_{N_c}(\lambda)\stackrel{\otimes}{,}T_{N_c}(\lambda')\right\}
= \left[ r(\lambda - \lambda'), T_{N_c}(\lambda)\otimes
T_{N_c}(\lambda')\right]
\ee
and the integrals of motion of the Toda chain are generated
by another form of spectral equation
\be\label{specTC0}
\det_{2\times 2}\left( T^{TC}_{N_c}(\lambda )
- w\right) = w^2 - w\Tr T^{TC}_{N_c}(\lambda ) + \det T^{TC}_{N_c}(\lambda ) =
w^2 - w\Tr T^{TC}_{N_c}(\lambda ) + 1 = 0
\ee
or
\be\label{specTC}
{\cal P}(\lambda ,w) = \Tr T^{TC}_{N_c}(\lambda) - w - \frac{1}{w} =
2P_{N_c}(\lambda) - w - \frac{1}{w} = 0
\ee
(We used the fact that
$\det_{2\times 2} L^{TC}(\lambda) = 1$ leads to $\det_{2\times 2} T_{N_c}^{TC}
(\lambda) = 1$.) The r.h.s. of (\ref{specTC}) is a polynomial of degree $N_c$
in $\lambda$, with the coefficients being the integrals of motion since
\be\label{trcom}
\left\{ \Tr T_{N_c}(\lambda ), \Tr T_{N_c}(\lambda' )\right\} =
\Tr \left\{ T_{N_c}(\lambda )\stackrel{\otimes}{,}T_{N_c}(\lambda' )\right\} =
\nn \\
= \Tr \left[ r(\lambda - \lambda' ), T_{N_c}(\lambda )\otimes
T_{N_c}(\lambda' )\right] = 0
\ee
For the particular choice of $L$-matrix (\ref{LTC}), the inhomogenities
of the chain, $\lambda_i$, can be absorbed into the redefinition
of the momenta $p_i \rightarrow p_i - \lambda_i$.

In what follows we consider possible elliptic deformations of
two Lax representations of the Toda chain. The deformation of the
$N_c\times N_c$ representation provides the Calogero-Moser model
while the deformation of the spin-chain $2\times 2$ representation gives rise
to the Sklyanin $XYZ$ model.

In addition to the curve (\ref{SpeC}), (\ref{fsc-Toda}), (\ref{hypelTC}) and
(\ref{specTC}) to define the Toda-\-chain system one needs a generating
1-differential $dS^{TC}$. I will discuss it in detail later now only mentioning
that $dS^{TC}= \lambda {dw\over w}$ in the Toda chain
case has exactly
the form of (\ref{dS}) where $\lambda $ in this example is a hyperelliptic
co-ordinate in (\ref{fsc-Toda}) and (\ref{hypelTC}). Its periods
\be\label{perTC}
{\bf a} = \oint _{\bf A}dS^{TC} = \oint _{\bf A}\lambda {dw\over w}
\ \ \ \ \ \ \ \
{\bf a}_D = \oint _{\bf B}dS^{TC} = \oint _{\bf B}\lambda {dw\over w}
\ee
define the BPS massive spectrum and prepotential of the pure ${\cal N}=2$
SUSY Yang-\-Mills theory.

\subsection{Elliptic deformation of the $N_c\times N_c$ representation:
the Calogero-Moser model \label{Cal}}

The $N_c\times N_c$ matrix Lax operator for the $GL(N_c)$ Calogero system is
\cite{KriCal}
\be\label{LaxCal}
{\cal L}^{Cal}(\xi) =
\left({\bf pH} + \sum_{\balpha}F({\bf {\goth q}\balpha}|\xi)
E_{\balpha}\right) = \nn \\
= \left(\begin{array}{cccc}
 p_1 & F({\goth q}_1-{\goth q}_2|\xi) & \ldots &
F({\goth q}_1 - {\goth q}_{N_c}|\xi)\\
F({\goth q}_2-{\goth q}_1|\xi) & p_2 & \ldots &
F({\goth q}_2-{\goth q}_{N_c}|\xi)\\
 & & \ldots  & \\
F({\goth q}_{N_c}-{\goth q}_1|\xi) &
F({\goth q}_{N_c}-{\goth q}_2|\xi)& \ldots &p_{N_c}
\end{array} \right)
\ee
The function $F({\goth q}|\xi) = \frac{g}{\omega}
\frac{\sigma({\goth q}+\xi)}{\sigma({\goth q})\sigma(\xi)}
e^{\zeta({\goth q})\xi}$ is expressed in terms of the Weierstrass elliptic
functions and, thus, the Lax operator ${\cal L}(\xi)$ is
defined on the elliptic curve $E(\tau)$ (complex torus
with periods $\omega, \omega '$ and modulus $\tau = \frac{\omega '}{\omega}$).
The Calogero coupling constant is $\frac{g^2}{\omega^2} \sim m^2$,
where in the $4d$
interpretation $m$ plays the role of the mass of the adjoint matter
${\cal N}=2$ hypermultiplet breaking ${\cal N}=4$ SUSY down to ${\cal N}=2$
\cite{WiDo}.

The spectral curve $\Sigma^{Cal}$ for the $GL(N_c)$ Calogero system is
given by:
\be\label{fscCal}
\det_{N_c\times N_c} \left({\cal L}^{Cal}(\xi) - \lambda\right) = 0
\ee
The BPS masses ${\bf a}$ and ${\bf a}_D$ are now the periods of the
generating 1-differential
\be\label{dSCal}
dS^{Cal} \cong \lambda d\xi
\ee
along the non-contractable contours on $\Sigma^{Cal}$.
Integrability of the Calogero system is implied by the Poisson
structure of the form
\be\label{lin-r}
\left\{ {\cal L}(\xi)\stackrel{\otimes}{,}{\cal L}(\xi ')\right\}
= \left[ {\cal R}_{12}^{Cal}(\xi ,\xi '),\ {\cal L}(\xi)\otimes {\bf 1}\right]
-
\left[ {\cal R}_{21}^{Cal}(\xi ,\xi '),\ {\bf 1} \otimes {\cal L}(\xi') \right]
\ee
with the {\it dynamical} elliptic ${\cal R}$-matrix \cite{Rmat},
guaranteeing that the eigenvalues of the matrix ${\cal L}$ are in involution.

In order to recover the Toda-chain system, one takes the double-scaling
limit \cite{Ino}, when $g \sim m$ and $-i\tau$ both go
to infinity (and
${\goth q}_i-{\goth q}_j={\2}\left[(i-j)\log g +(q_i-q_j)\right]$)
so that the dimensionless coupling $\tau$ gets
substituted by a dimensional parameter $\Lambda^{N_c} \sim
m^{N_c}e^{i\pi\tau}$. In this limit,
the elliptic curve $E(\tau)$ degenerates into the (two-punctured) Riemann
sphere with coordinate $w = e^{\xi }e^{i\pi\tau}$ so that
\be
dS^{Cal} \rightarrow dS^{TC} \cong \lambda\frac{dw}{w}
\ee
The Lax operator of the Calogero system turns into that of the
$N_c$-periodic Toda chain (\ref{LaxTC}):
\be
{\cal L}^{Cal}(\xi )d\xi \rightarrow {\cal L}^{TC}(w)\frac{dw}{w}
\ee
and the spectral curve acquires the form of (\ref{SpeC}).
In contrast to the Toda case, (\ref{fscCal}) can {\it not} be rewritten in
the form (\ref{fsc-Toda}) and specific $w$-dependence of the spectral
equation (\ref{SpeC}) is not preserved by embedding of Toda into Calogero-Moser
particle
system. However, the form (\ref{fsc-Toda}) can be naturally preserved by
the alternative deformation of the Toda-chain system when it
is considered as (a particular case of) a spin-chain model.

To deal with the ``elliptic" deformations of the Toda chain below, we will use
a non-standard normalization of the Weierstrass
$\wp $-function defined by
\be\label{wpmod}
\wp(\xi |\tau) = \sum_{m,n = -\infty}^{+\infty}
\frac{1}{(\xi + m + n\tau)^2} - {\sum_{m,n=-\infty}^{+\infty}}'
\frac{1}{(m+n\tau)^2}
\ee
so that it is double periodic in $\xi $ with periods $1$ and
$\tau = {\omega '\over \omega}$ (that differs from a standard definition
by a factor of $\omega^{-2}$ and by
rescaling $\xi \rightarrow \omega\xi$). According to (\ref{wpmod}), the
values of $\wp(\xi |\tau)$ in the half-periods, $e_a = e_a(\tau )$, $a=1,2,3$,
are also the functions only of $\tau$ -- again differing by a factor
of $\omega^{-2}$ from the conventional definition.
The complex torus $E(\tau )$ can be defined as ${\bf C}/{\bf Z}\oplus
\tau{\bf Z}$
with a
``flat" co-ordinate $\xi$ defined $modulo(1,\tau)$.
Alternatively, any torus (with a marked point) can be described as
elliptic curve
\be\label{ell}
y^2 = (x -  e_1)(x - e_2)(x - e_3) \ \ \ \ \
x
= \wp(\xi ) \ \ \  \ \ y = - \frac{1}{2}\wp'(\xi )
\ee
and the canonical holomorphic 1-differential is
\be
d\xi = 2\frac{dx}{y}
\ee
In the simplest example of $N_c=2$,
the spectral curve $\Sigma^{Cal}$ has genus 2. Indeed,
in this particular case, eq.(\ref{fscCal}) turns into
\be\label{caln2}
{\cal P}(\lambda;x,y) = \lambda ^2 - h_2 + \frac{g^2}{\omega^2} x = 0
\ee
This equation says that with any value of $x$ one
associates two points of $\Sigma^{Cal}$,
$\lambda = \pm\sqrt{h_2 - \frac{g^2}{\omega^2}x}$, i.e.
it describes $\Sigma^{cal}$ as a double covering of the
elliptic curve $E(\tau)$ ramified at the points
$x = \left( {\omega\over g}\right)^2h_2$
and $x = \infty$. In fact, since $x$ is an elliptic
coordinate on $E(\tau)$ (when elliptic curve itself is
treated as a double covering over the Riemann sphere $CP^1$),
$x = \left( {\omega\over g}\right)^2h_2$ corresponds to a {\it pair} of
points on $E(\tau)$ distinguished by the sign of $y$. This would be true
for $x = \infty$
as well, but $x = \infty$ is one of the branch points in
our parameterization (\ref{ell}) of $E(\tau)$. Thus, the {\it two} cuts
between $x = \left( {\omega\over g}\right)^2h_2$ and $x=\infty$ on every
sheet of
$E(\tau)$ touching at the common end at $x=\infty$ become effectively
a {\it single} cut between $\left(\left( {\omega\over g}\right)^2h_2, +\right)$
and $\left(\left( {\omega\over g}\right)^2h_2, -\right)$. Therefore, we can
consider the spectral
curve $\Sigma^{Cal}$ as two tori $E(\tau)$ glued along one cut, i.e.
$\Sigma^{Cal}_{N_c=2}$ has genus 2.
It turns out to be a hyperelliptic curve (for $N_c = 2$ only!)
after substituting in (\ref{caln2}) $x$ from the second equation to the
first one.

Two holomorphic 1-differentials on $\Sigma^{Cal}$ can be chosen to be
\be\label{holn2}
v = \frac{dx}{y} \sim \frac{\lambda d\lambda}{y}
\ \ \ \ \ \ \
V =  \frac{dx}{y\lambda}\sim \frac{d\lambda }{y}
\ee
so that
\be
dS \cong \lambda d\xi =
\sqrt{h_2 - \frac{g^2}{\omega^2}\wp(\xi )}d\xi =
\frac{dx}{y}\sqrt{h_2 - \frac{g^2}{\omega^2}x}
\ee
It is easy to check the basic property (\ref{hol}):
\be
\frac{\partial dS}{\partial h_2} \cong \frac{1}{2}\frac{dx}{y\lambda}
\ee
The fact that only one of two holomorphic 1-differentials (\ref{holn2})
appears at the r.h.s. is related to their different parity
with respect to the ${\bf Z}_2\otimes {\bf Z}_2$ symmetry of $\Sigma^{Cal}$:
$y \rightarrow -y$ and $\lambda \rightarrow -\lambda$.
Since $dS$ has certain parity, its
integrals along the two of four elementary non-contractable cycles
on $\Sigma^{Cal}$ automatically vanish leaving only two
non-vanishing quantities $a$ and $a_D$, as necessary for
the $4d$ interpretation. Moreover, these two
non-vanishing
integrals can be actually evaluated in terms of the ``reduced"
genus-{\it one} curve
\be
Y^2 = (y\lambda)^2 = \left(h_2 - \frac{g^2}{\omega^2}x\right)
\prod_{a=1}^3 (x - e_a),
\ee
with $dS \cong \left(h_2 - \frac{g^2}{\omega^2}x\right)\frac{dx}{Y}$.
Since now $x = \infty$ is no longer a ramification point, $dS$
obviously has simple poles at $x = \infty$ (at two points on the
two sheets of $\Sigma^{Cal}_{reduced}$) with the residues
$\pm\frac{g}{\omega} \sim \pm m$.

The opposite limit of the Calogero-Moser system with vanishing coupling
constant $g^2 \sim m^2 \rightarrow 0$ corresponds to the ${\cal N}=4$ SUSY
Yang-Mills theory with identically vanishing $\beta $-function.
The corresponding integrable
system is a collection of {\it free} particles and the generating differential
$dS \cong \sqrt{h_2}\cdot d\xi $ is just a {\it holomorphic} differential on
$E(\tau )$.

\subsection{From Toda to Spin Chains}

Now, let us discuss another deformation of the Toda chain corresponding to
the coupling of the ${\cal N}=2$ SYM theory to the fundamental matter.
According to \cite{SW2}, \cite{fumat}, the spectral curves for
the ${\cal N}=2$ SQCD with any $N_f < 2N_c$ have the same
form as (\ref{spec}) with a less trivial monodromy matrix
\be\label{trdet}
\Tr\ T_{N_c}(\lambda) = P_{N_c}(\lambda) + R_{N_c-1}(\lambda),
\ \ \ \ \det T_{N_c}(\lambda ) = Q_{N_f}(\lambda ),
\ee
and $Q_{N_f}(\lambda)$ and $R_{N_c-1}(\lambda)$ are certain
$h$-{\it independent} polynomials of $\lambda$ (remind that for the Toda chain
with the Lax matrix (\ref{LTC})
$\det _{2\times 2} T_{N_c}^{\rm TC}(\lambda) =
\prod_{i=1}^{N_c} \det _{2\times 2} L_i^{\rm TC}(\lambda-\lambda_i) = 1$
and
$\Tr\  T_{N_c}^{\rm TC}(\lambda ) = P_{N_c}(\lambda )$).
Two descriptions, (\ref{SpeC}) and (\ref{specTC0}),
are identically equivalent for the Toda
chain, but their {\it deformations} are very different:
the ``chain" representation (\ref{specTC0}), (\ref{specTC})
is naturally embedded into the family of
$XYZ$ spin chains \cite{FT,Skl}.

Our proposal is to look at the orthogonal to the previous generalization of
the Toda chain,
i.e. deform eqs.(\ref{quadrP})-(\ref{trcom}) preserving the Poisson brackets
\be\label{quadr-r}
\left\{L(\lambda)\stackrel{\otimes}{,}L(\lambda')\right\} =
\left[ r(\lambda-\lambda'),\ L(\lambda)\otimes L(\lambda')\right],
\nn \\
\left\{T_{N_c}(\lambda)\stackrel{\otimes}{,}T_{N_c}(\lambda')\right\} =
\left[ r(\lambda-\lambda'),\
T_{N_c}(\lambda)\otimes T_{N_c}(\lambda')\right],
\ee
and, thus, the possibility to build a
monodromy matrix $T(\lambda)$ by multiplication of $L_i(\lambda)$'s.
The full spectral curve for the periodic {\it inhomogeneous} spin chain is
given by:
\be\label{fsc-SCh}
\det\left(T_{N_c}(\lambda) -  w\right) = 0,
\ee
with the inhomogeneous $T$-matrix still defined by (\ref{monomat}),
satisfying (\ref{quadr-r}), while generating 1-form is now
\be\label{1f}
dS = \lambda\frac{dW}{W}
\ee
In the case of $sl(2)$ spin chains, the spectral
equation can be written as
\be\label{fsc-sc1}
w + \frac{\det_{2\times 2} T_{N_c}(\lambda)}{w} =
 \Tr_{2\times 2}T_{N_c}(\lambda ),
\ee
or
\be
W + \frac{1}{W} = \frac{\Tr_{2\times 2}T_{N_c}(\lambda)}
{\sqrt{\det_{2\times 2} T_{N_c}(\lambda)}}.
\label{fsc-sc2}
\ee
The r.h.s. of this equation contains the dynamical variables of
the spin system only in the special combinations -- its
Hamiltonians (which are all in involution, i.e. Poisson-commuting).
It is this peculiar shape (quadratic $w$-dependence) that suggests the
identification of the periodic $sl(2)$ spin chains with solutions to the
SW problem with the fundamental matter supermultiplets.

\subsection{$XXX$ Spin Chain and the Low Energy SYM with $N_f < 2N_c$}

The $2\times 2$ Lax matrix for the $sl(2)$ $XXX$ chain is
\be\label{laxrat}
L(\lambda) = \lambda \cdot {\bf 1} + \sum_{a=1}^3 S_a\cdot\sigma^a.
\ee
The Poisson brackets of the dynamical variables $S_a$, $a=1,2,3$
(taking values in the algebra of functions)
are implied by (\ref{quadr-r}) with the rational $r$-matrix
\be\label{rat-r}
r(\lambda) = \frac{1}{\lambda}\sum_{a=1}^3 \sigma^a\otimes \sigma^a.
\ee
In the $sl(2)$ case, they are just
\be\label{Scomrel}
\{S_a,S_b\} = i\epsilon_{abc} S_c,
\ee
i.e. $\{S_a\}$ plays the role of angular momentum (``classical spin'')
giving the name ``spin-chains'' to the whole class of systems.
Algebra (\ref{Scomrel}) has an obvious Casimir operator
(an invariant, Poisson-\-commuting with all the generators $S_a$),
\be\label{Cas}
K^2 = {\bf S}^2 = \sum_{a=1}^3 S_aS_a,
\ee
so that
\be\label{detTxxx}
\det_{2\times 2} L(\lambda) = \lambda^2 - K^2,
\nn \\
\det_{2\times 2} T_{N_c}(\lambda) = \prod_{i=N_c}^1
\det_{2\times 2} L_i(\lambda-\lambda_i) =
\prod_{i=N_c}^1 \left((\lambda - \lambda _i)^2 - K_i^2\right) = \nn \\
= \prod_{i=N_c}^1(\lambda + m_i^+)(\lambda + m_i^-)
= Q_{2N_c}(\lambda),
\ee
where we assumed that the values of spin $K$ can be different at
different nodes of the chain, and
\footnote{
Eq.(\ref{mpm}) implies that the limit of vanishing masses, all
$m_i^\pm = 0$, is associated with the {\it homogeneous} chain
(all $\lambda_i = 0$) and vanishing spins at each site (all $K_i = 0$).
It deserves noting that a similar situation was considered by
L.Lipatov \cite{L} in the study of {\it the high}-energy limit of
the ordinary (non-supersymmetric) QCD.
The spectral equation is then the classical limit of the
Baxter equation from \cite{FK}.  }
\be
m_i^{\pm} = -\lambda_i \mp K_i.
\label{mpm}
\ee
While the determinant of monodromy matrix (\ref{detTxxx})
depends on dynamical variables
only through the Casimir values $K_i$ of the Poisson algebra, the dependence of
the trace
${\cal T}_{N_c}(\lambda) =
\frac{1}{2}\Tr_{2\times 2}T_{N_c}(\lambda)$ is less trivial.
Still, as usual for integrable systems, it depends
on $S_a^{(i)}$ only through the Hamiltonians of the spin chain (which are not
Casimirs but Poisson-commute with {\it each other}).

In order to get some impression how the Hamiltonians
look like, we present explicit examples of monodromy matrices
for $N_c = 2$ and $3$. Hamiltonians depend non-\-trivially
on the $\lambda_i$-parameters (in\-homo\-geneities of the chain)
and the coefficients
in the spectral equation (\ref{fsc-SCh}) depend only on the
Hamiltonians and symmetric functions of the $m$-parameters (\ref{mpm}),
i.e. the dependence of $\{\lambda_i\}$ and $\{K_i\}$ is rather special.
This property is crucial for identification of the $m$-parameters
with the masses of the matter supermultiplets in the ${\cal N}=2$ SQCD.
The explicit examples can be found in \cite{Spin}.

\subsection{N$_f$ = 2N$_c$: Generic spin-chain models and Sklyanin algebra}
However,
the story can not be complete without studying the most
intriguing ``elliptic'' case of $N_f=2N_c$, when the $4d$ theory
is supposed to be UV-finite (at least at some particular values of moduli) and
then possesses an extra {\it dimensionless}
parameter: the UV non-abelian coupling constant
$\tau = \frac{8\pi i}{e^2} + \frac{\theta}{\pi}$.

The most general theory of this sort is known as Sklyanin $XYZ$ spin chain
with the elementary $L$-operator defined on the elliptic curve
$E(\tau)$ and is explicitly given by (see \cite{FT} and references therein):
\be\label{39}
L^{Skl}(\xi) = S^0{\bf 1} + i\frac{g}{\omega}\sum_{a=1}^3 W_a(\xi)S^a\sigma_a
\ee
where
\be
W_a(\xi) = \sqrt{e_a - \wp\left({\xi}|\tau\right)} =
i\frac{\theta'_{11}(0)\theta_{a+1}\left({\xi}\right)}{\theta_{a+1}(0)
\theta_{11}\left({\xi}\right)}\\
\theta_2\equiv\theta_{01},\ \ \ \theta_3\equiv\theta_{00},
\ \ \ \theta_{4}\equiv\theta_{10}
\ee
Let us note that our spectral parameter $\xi$ is connected with the standard
one $u$ \cite{FT} by the relation $u=2K\xi$, where
$K\equiv\int_0^{{\pi\over 2}}{dt\over\sqrt{1-k^2\sin^2t}}={\pi\over
2}\theta_{00}^2(0)$, $k^2\equiv{e_1-e_2\over e_1-e_3}$
so that $K\to{\pi\over 2}$ as $q\to 0$. This factor results into additional
multiplier $\pi$ in the trigonometric functions in the limiting cases below.

The Lax operator (\ref{39})
satisfies the Poisson relation (\ref{quadrP}) with the
numerical {\it elliptic} $r$-matrix $r(\xi)={i{g\over\omega}}\sum_{a=1}^3
W_a(\xi)\sigma_a\otimes\sigma_a$, which
implies that $S^0, S^a$ form the (classical)
Sklyanin algebra \cite{Skl1,Skl}:
\be\label{sklyal}
\left\{S^a, S^0\right\} = 2i\left(\frac{g}{\omega}\right)^2
\left(e_b - e_c\right)S^bS^c
\nn \\
\left\{S^a, S^b\right\} = 2iS^0S^c
\ee
with the obvious notation: $abc$ is the triple $123$ or its cyclic
permutations.

The coupling constant  ${g\over \omega}$ can be eliminated by simultaneous
rescaling of the $S$-variables and the symplectic form:
\be
S^a = \frac{\omega}{g}\hat S^a \ \ \ \
S^0 = \hat S^0\ \ \ \
\{\ , \ \} \rightarrow \frac{g}{\omega}\{\ ,\ \}
\ee
Then
\be
L(\xi) = \hat S^0 {\bf 1} + i\sum_{a=1}^3 W_a(\xi)\hat S^a\sigma_a
\ee
\be\label{sklyaln}
\left\{ \hat S^a,\ \hat S^0\right\} = 2i
\left(e_b - e_c\right) \hat S^b\hat S^c
\ \ \ \ \ \
\left\{ \hat S^a,\ \hat S^b\right\} = 2i\hat S^0\hat S^c
\ee
In the rational limit both $\omega , \omega '\rightarrow\infty $, then
(\ref{sklyal}) turns into
\be\label{ratsklya}
\{ S^a,S^0\} = 0
\nn \\
\{ S^a,S^b\} = 2i\epsilon ^{abc}S^0S^c
\ee
i.e. $S^0$ itself becomes a Casimir operator (constant), while the remaining
$S^a$ form a classical angular-momentum (spin) vector (\ref{Scomrel}).
The corresponding Lax operator (\ref{laxrat})
$L \equiv \lambda L_{XXX} = \lambda {\bf 1}+ {\bf S\cdot\bsigma}$
describes the $XXX$ spin chain with the rational $r$-matrix (\ref{rat-r}).

The determinant
$\det _{2\times 2} \hat L(\xi)$ is equal to
\be
\det _{2\times 2} \hat L(\xi) = \hat S_0^2 + \sum_{a=1}^3 e_a\hat S_a^2
- \wp(\xi)\sum_{a=1}^3\hat S_a^2 = K - M^2\wp(\xi) =  K - M^2 x
\ee
where
\be\label{casi}
K = \hat S_0^2 + \sum_{a=1}^3 e_a(\tau)\hat S_a^2 \ \ \
\ \ \ \
M^2 =  \sum_{a=1}^3 \hat S_a^2
\ee
are the Casimir operators of the Sklyanin algebra (i.e. Poisson
commuting
with all the generators $\hat S^0$, $\hat S^1$, $\hat S^2$,
$\hat S^3$). The determinant of the monodromy matrix (\ref{monomat})
is
\be\label{54}
Q(\xi) = \det _{2\times 2} T_{N_c}(\xi) =
\prod_{i=1}^{N_c} \det _{2\times 2} \hat L(\xi - \xi_i) =
\prod_{i=1}^{N_c} \left( K_i - M^2_i\wp(\xi - \xi_i)\right)
\ee
while the trace $P(\xi) = \frac{1}{2}\Tr T_{N_c}(\xi)$
generates mutually Poisson-commuting Hamiltonians, since
\be
\left\{\Tr T_{N_c}(\xi),\
\Tr T_{N_c}(\xi') \right\} = 0
\ee
For example, in the case of the {\it homogeneous} chain (all $\xi_i = 0$
in (\ref{54}))
$\Tr T_{N_c}(\xi)$ is a combination of the
polynomials:
\be
P(\xi) = Pol^{(1)}_{\left[\frac{N_c}{2}\right]}(x) +
y Pol^{(2)}_{\left[\frac{N_c-3}{2}\right]}(x),
\ee
where $\left[\frac{N_c}{2}\right]$ is integral part
of ${N_c\over 2}$, and the coefficients of $Pol^{(1)}$ and $Pol^{(2)}$
are Hamiltonians of the $XYZ$ model
\footnote{For the {\it inhomogeneous} chain the explicit expression for the
trace is more sophisticated: one should make use of the
formulas like
$$
\wp(\xi - \xi_i) = \left(\frac{\wp'(\xi) + \wp'(\xi_i)}
{\wp(\xi) - \wp(\xi_i)}\right)^2 - \wp(\xi) - \wp(\xi_i) =
4\left(\frac{y+y_i}{x-x_i}\right)^2 - x - x_i
$$
}.
As a result, the spectral equation (\ref{fsc-sc2}) for the $XYZ$ model
acquires the form:
\be\label{specXYZ}
w + \frac{Q(\xi)}{w} = 2P(\xi),
\ee
where for the {\it homogeneous} chain $P$ and $Q$
are polynomials in $x = \wp(\xi)$
and $y =   \frac{1}{2}\wp'(\xi)$.
Eq. (\ref{specXYZ}) describes the double covering of the elliptic
curve $E(\tau)$:
with generic point $\xi \in E(\tau)$ one associates the
two points of $\Sigma^{XYZ}$, labeled by two roots $w_\pm$
of equation (\ref{specXYZ}).
The ramification points correspond to
$w_+ = w_- = \pm\sqrt{Q}$, or
$Y = \frac{1}{2}\left(w - \frac{Q}{w}\right) =
\sqrt{P^2 - Q}= 0$.

The curve (\ref{specXYZ}) is in fact similar to that of $N_c=2$
Calogero-Moser system (\ref{caln2}).
The difference is that now $x = \infty$ is {\it not} a branch point,
therefore, the number of cuts on the both copies of $E(\tau)$ is $N_c$ and the
genus of the spectral curve is $N_c+1$.
Rewriting analytically $\Sigma^{XYZ}$ as a system of equations
\be\label{cxyz}
y^2 = \prod_{a=1}^3 (x - e_a), \nn \\
Y^2 = P^2 - Q
\ee\label{hdb}
the set of holomorphic 1-differentials on $\Sigma^{XYZ}$ can be chosen as
\be
v = \frac{dx}{y},\ \ \ \ \
V_\alpha = \frac{x^\alpha dx}{yY} \ \ \
\alpha = 0,\ldots,
\left[\frac{N_c}{2}\right] \ \ \ \ \ \
\tilde V_\beta = \frac{x^\beta dx}{Y} \ \ \
\beta = 0, \ldots,
\left[\frac{N_c-3}{2}\right]
\ee
with the total number of holomorphic 1-differentials
$1 + \left(\left[\frac{N_c}{2}\right] + 1\right) +
\left(\left[\frac{N_c-3}{2}\right] + 1\right) = N_c+1$
being equal to the genus of $\Sigma^{XYZ}$.

Now, given the spectral curve one can try to
write down the ``generating" 1-differential
$dS$ which obeys the basic defining property
(\ref{hol}). For the Toda chain it can be chosen in two different ways
\be\label{dstc}
d\Sigma^{TC} \cong d\lambda\log w \ \ \ \ \  dS^{TC} \cong \lambda{dw\over w}
\ \ \ \ \ \
d\Sigma ^{TC} = - dS^{TC} + df^{TC}
\ee
Both $d\Sigma ^{TC}$ and $dS^{TC}$ obey the basic property (\ref{hol}) and,
while $f^{TC}$ itself is {\it not} its variation, $\delta f^{TC} =
\lambda{\delta w\over w}$ appears to be a (meromorphic) single-valued
function on $\Sigma^{TC}$.
In the $XXX$ case \cite{Spin}, one has almost the same formulas as
(\ref{dstc})
\be\label{dsxxx}
d\Sigma^{XXX} \cong d\lambda\log W \ \ \ \ \
dS^{XXX} \cong \lambda{dW\over W} \ \ \ \ \ \ \
d\Sigma ^{XXX} = - dS^{XXX} +
df^{XXX} \ \ \ \ \
W \equiv {w\over\sqrt{\det T_{N_c}(\lambda )}}
\ee
For the
$XYZ$ model (\ref{specXYZ}) the generating 1-form(s) $dS^{XYZ}$ can be
defined as
\be\label{dsxyz}
d\Sigma ^{XYZ} \cong d\xi \cdot \log W
\ \ \ \ \ \
dS^{XYZ} \cong \xi {dW\over W} = - d\Sigma ^{XYZ} + d(\xi\log W)
\ee
Now, under the variation of moduli (which are all contained in $P$,
while $Q$ is moduli independent),
\be
\delta(d\Sigma ^{XYZ}) \cong \frac{\delta W}{W}d\xi =
\frac{\delta P(\xi)}{\sqrt{P(\xi)^2-Q(\xi)}} d\xi
= \frac{dx}{yY}\delta P
\ee
and, according to (\ref{hdb}), the r.h.s. is a {\it holomorphic}
1-differential on the spectral curve (\ref{specXYZ}).
The singularities of $d\Sigma ^{XYZ}$ are located at the points where $W=0$ or
$W=\infty$, i.e.  at zeroes of $Q(\xi)$ or poles of $P(\xi)$.
In the vicinity of a singular point, $d\Sigma ^{XYZ}$ is not
single-valued but acquires addition $2\pi id\xi$ when circling around this
point.
The difference between $d\Sigma$ and $dS$ is again a total derivative,
but $\delta f^{XYZ} = \xi{\delta W\over W}$ is not a single-valued function.
In contrast to $d\Sigma ^{XYZ}$,
$dS^{XYZ}$ has simple poles at $W=0,\infty$ with the residues
$\left.\xi\right|_{w=0,\infty}$,
which are defined modulo $1,\tau$. Moreover, $dS^{XYZ}$ itself is
multivalued: it changes by $(1,\tau)\times\frac{dW}{W}$ when circling
along non-contractable cycles on $E(\tau)$.

Naively, neither $d\Sigma ^{XYZ}$ nor $dS^{XYZ}$ can naively play the role of
the Seiberg-Witten
1-form -- which is believed to possess well-defined residues, interpreted as
masses of the matter hypermultiplets \cite{SW2}.

One can naturally assume that the $XYZ$ chain, which is an
elliptic deformation of the $XXX$ chain known
to describe ${\cal N}=2$  supersymmetric QCD with $N_f < 2N_c$,
can be associated with the  $N_f = 2N_c$ case.
This would provide a description of the conformal (UV-finite)
supersymmetric QCD, differing from the conventional one \cite{SW2,fumat}.
However, as demonstrated above,
there are several serious differences between $XYZ$
and $XXX$ models, which should be kept in mind.

{\bf 1)} Normally, there are two natural ways to introduce
generating 1-form -- these are
represented by the meromorphic 1-differentials $dS$ and $d\Sigma$
in the main text.
Usually, {\it both} satisfy (\ref{hol}); $d\Sigma$
has no simple poles, but is not single-valued on $\Sigma$,
while $dS$ is single-valued and possesses simple poles
(of course, in general $dS \ncong d\Sigma$). The proper generating
1-form is $dS$.
However, in the $XYZ$ case, {\it both} $dS$ and $d\Sigma$ are {\it not}
single-valued. Moreover, the residues of $dS$ -- identified
with masses of the matter hypermultiplets in the framework of \cite{SW2}
-- are defined only modulo $(1,\tau)$.

{\bf 2)} As $\tau \rightarrow i\infty$, the $XYZ$ model turns into
$XXZ$ rather than $XXX$ one. This makes the description of the
``dimensional transmutation'' regime rather tricky.

{\bf 3)} Starting from the spectral curve (\ref{SpeC}) for the
Toda-chain (pure ${\cal N}=2$ SYM), the Calogero-Moser deformation is
associated with ``elliptization" of the $w$-variable, while
the $XYZ$-deformation -- with that of the $\lambda$-variable.
It is again nontrivial to reformulate the theory in such
a way that the both deformations become of the same nature.
One of the most naive pictures would associate the Hitchin-type
(Calogero) models with the ``insertion" of an $SL(N_c)$-orbit at
one puncture on the elliptic curve, while the
spin-chain ($XYZ$) models -- with the $SL(2)$-orbits
at $N_c$ punctures. Alternatively, one can say that the
$XYZ$-type ''elliptization", while looking local (i.e. $L$-operators
at every site are deformed independently), is in fact a global
one (all the $L$'s can be elliptized  only simultaneously, with
the same $r$-matrix and $\tau$), --  but this is not clearly
reflected in existing {\it formalism}, discussed in this section.
Moreover, the proper formalism should naturally allow one
to include any simple Lie group (not only $SL(N_c)$)
and any representations (not only adjoint or fundamental).

Already these comments are enough to demonstrate that the hottest
issues of integrability and quantum group theory (like notions
of elliptic groups and dynamical $R$-\-matrices) can be of immediate
importance for the Seiberg-Witten (and generic duality) theory.

\section{Symplectic structure}

Now let us turn to the discussion of a more subtle point -- why the
generating 1-form (\ref{dS}) indeed describes an integrable system.
To do this I will discuss the symplectic structure on the space of the
finite-gap solutions.
This symplectic structure was introduced in
\cite{DKN} and recently proposed in \cite{KriPho} as coming directly
from the symplectic form on the space of all the solutions to the hierarchy.
Below, a very simple and straightforward proof of this result is
presented \cite{M5} and the relation with the analogous object in
low-dimensional non-perturbative string theory \cite{KM} is discussed.

To prove that (\ref{dS}) is a generating one-form of the whole hierarchy
one starts with the variation of the generating function
\be\label{S}
S(\Sigma ,{\bf \gamma}) = \sum _i \int ^{\gamma _i}Edp
\ee
(where $dE$ and $dp$ ($= d\lambda $ and $={dw\over w}$) in the particular case
above) are two meromorphic
differentials on a spectral curve $\Sigma $ and $\bf \gamma $ is the divisor
of the solution (poles of the BA function))
\be\label{var}
\delta S = \sum _i (Edp)(\gamma _i) + \sum _i \int ^{\gamma _i}\delta E dp
\ \ \ \ \ \
\delta ^2 S = \delta\left( \sum _i (Edp)(\gamma _i)\right)
+ \sum _i (\delta E dp)(\gamma _i)
\ee
From $\delta ^2 S = 0$ it follows that
\be\label{wsform}
\varpi  = \delta E\wedge\delta p =
\delta\left( \sum _i (Edp)(\gamma _i)\right) =
- \sum _i (\delta E dp)(\gamma _i)
\ee
Now, the variation $\delta E$ (for constant $p$) follows from the Lax
equation (auxiliary linear problem)
\be\label{lax}
{\partial\over\partial t}\psi = {\cal L}\psi \ (= E\psi )
\ee
so that
\be\label{varE}
\delta E = {\langle \psi ^{\dagger}\delta {\cal L}\psi\rangle\over
\langle \psi ^{\dagger}\psi\rangle}
\ee
and one concludes that
\be\label{general}
\varpi =  - \langle \delta{\cal L}\sum _i \left( dp{\psi ^{\dagger}\psi\over
\langle \psi ^{\dagger}\psi\rangle}\right) (\gamma _i) \rangle
\ee
Let us turn to several important examples.\\
{\bf KP/KdV}. In the KP-case the equation (\ref{lax}) looks as
\be\label{laxkp}
{\partial\over\partial t}\psi = \left({\partial ^2\over\partial x^2} +
u\right) \psi \ (= E\psi )
\ee
therefore the equation (\ref{general}) implied by
$\langle \psi ^{\dagger}\psi\rangle = \int _{dx}\psi ^{\dagger}(x,P)\psi (x,P)$
and $\delta{\cal L} = \delta u(x)$ gives
\be\label{kpcase}
\varpi  =  - \int _{dx} \delta u(x) \sum _i
\left( {dp\over\langle \psi ^{\dagger}\psi\rangle}\psi ^{\dagger}(x)\psi (x)
\right) (\gamma _i)
\ee
The differential
$d\Omega = {dp\over\langle \psi ^{\dagger}\psi\rangle}
\psi ^{\dagger}(x)\psi (x)$ is
holomorphic on $\Sigma $ except for the ''infinity" point $P_0$ where it has
zero residue \cite{KriUMN}. Its variation
\footnote{It should be pointed out that the variation $\tilde\delta $
corresponds to a rather specific situation when one shifts only $\psi $
keeping $\psi ^{\dagger}$ fixed.}
\be\label{reskp}
{\tilde\delta}\left(\res _{P_0}d\Omega + \sum _i\res _{\gamma _i}d\Omega \right) = 0
\ee
can be rewritten as
\be\label{reskp2}
\delta v(x) + \sum _i d\Omega (\gamma _i) = 0
\ee
where $v(x)$ is a ''residue" of the BA function at the point $P_0$ obeying
$v'(x) = u(x)$. Substituting (\ref{reskp2}) into (\ref{general}) one gets
\be\label{kdv1}
\varpi = \int _{dx} \delta u(x) \int ^x _{dx'}\delta u(x')
\ee
or the {\it first} symplectic structure of the KdV equation.\\
{\bf Toda chain/lattice}. (The case directly related to the {\it
pure} SYM theory). One has
$$
\langle \psi ^{\dagger}\psi\rangle = \sum _n\psi ^+_n(P)\psi ^-_n(P)
$$
and the Lax equation acquires the form (\ref{laxtoda})
where $t = t_+ + t_-$ and $t_1 = t_+ - t_-$ is the first time of the Toda
chain, so that
\be\label{varEtoda}
\delta\lambda = {\sum _k \psi ^+_k \delta
p_k\psi ^-_k\over \langle \psi ^+\psi ^-\rangle}
\ee
and (\ref{general})
becomes
\be\label{toda}
\varpi  =  - \sum _k \delta p_k \sum _i \left(
{dp\over\langle \psi ^+\psi ^-\rangle}\psi ^+_k \psi ^-_k
\right) (\gamma _i)
\ee
and to get
\be\label{ochain}
\varpi = \sum _k \delta p_k\wedge\delta q_k
\ee
one has to prove
\be\label{varx}
\sum _i \left(
{dp\over\langle \psi ^+\psi ^-\rangle}\psi ^+_k \psi ^-_k
\right) (\gamma _i)  = \delta q_k
\ee
To do this one considers again
\be\label{restoda}
{\tilde\delta} \left(\res _{P_+} + \res _{P_-} +
\sum _i\res _{\gamma _i}\right)d\Omega _n = 0
\ \ \ \ \ \
d\Omega _n =
{dp\over\langle \psi ^+\psi ^-\rangle}\psi ^+_n \psi ^-_n
\ee
where the first two terms for $\psi ^{\pm}_n \stackreb{\lambda
\rightarrow\lambda (P_{\pm})}{\sim} e^{\pm q_n}\lambda ^{\pm n}(1 +
{\cal O}(\lambda ^{-1}))$ satisfying two ''shifted" equations (\ref{laxtoda})
(with
${\tilde q}_n$ and $q_n $ correspondingly) give $\delta q_n = {\tilde q}_n -
q_n$ while the rest -- the l.h.s. of (\ref{varx}).

{\bf Calogero-Moser system}.
Introducing the ''standard" $dE$ and $dp$ one the curve $\Sigma $
(\ref{fscCal})
with the 1-form (\ref{dSCal})
where $dp=d\xi$ is holomorphic on torus $\oint _A dp = \omega$,
$\oint _B dp = \omega '$ and $E=\lambda $
has $n-1$ poles with $residue = 1$ and $1$ pole with $residue = -(n-1)$,
the BA function is defined by \cite{KriCal}
\be\label{bacal}
{\cal L}^{Cal}(\xi ){\bf a} =  \lambda {\bf a}
\ee
with the essential singularities
\be\label{baprop}
a_i \stackreb{E = E_+}{\sim} e^{x_i\zeta (\xi )}\left( 1 + {\cal O}(\xi)\right)
\ \ \ \ \
a_i \stackreb{E \neq E_+}{\sim} e^{x_i\zeta (\xi)}\left( -{1\over n-1} +
{\cal O}(\xi)\right)
\ee
and (independent of dynamical variables) poles ${\bf \gamma}$.  Hence,
similarly to the above case for the eq. (\ref{general}) one has
$\langle \psi ^{\dagger}\psi\rangle = \sum _ia_i^{\dagger}(P) a_i(P)$,
$\delta {\cal L}^{Cal} = {\sum _ia_i^{\dagger}(P) \delta p_i a_i(P)\over
\sum _ia_i^{\dagger}(P) a_i(P)}$ so that
\be\label{calo}
\varpi =  - \sum _k \delta p_k \sum _i \left(
{d\xi\over\langle a^{\dagger}a\rangle}a^{\dagger}_k a_k
\right) (\gamma _i)
\ee
and the residue formula
\be\label{rescal}
{\tilde\delta} \left(\sum _{P_j:p = 0}\res _{P_j} +
\sum _i\res _{\gamma _i}\right)d\Omega _k = 0
\ \ \ \ \ \ \
d\Omega _k =
{d\xi\over\langle a^{\dagger}a\rangle}a^{\dagger}_k a_k
\ee
where the first sum is over all ''infinities" $p = 0$ at each sheet of the
cover (\ref{fscCal}). After variation and using (\ref{baprop}) it gives
again
\be\label{calogero}
\varpi = \sum _k \delta p_k\wedge \delta x_k
\ee
The general proof of the more cumbersome analog of the above derivation
can be found in \cite{KriPho}. To show how the above
formulas
work explicitly, let us, finally, demonstrate the existence of (\ref{reskp}),
(\ref{restoda}) and (\ref{rescal}) for the 1-gap solution. Let
\be\label{psya}
\psi = e^{x\zeta (z)}
{\sigma (x-z+\kappa)\over\sigma (x+\kappa)\sigma (z-\kappa)}
\ \ \ \  \ \ \ \
\psi ^{\dagger}= e^{-x\zeta (z)}
{\sigma (x+z+\kappa)\over\sigma (x+\kappa)\sigma (z+\kappa)}
\ee
be solutions to
\be\label{1gap}
(\partial ^2 + u)\psi = (\partial ^2 - 2\wp (x+\kappa))\psi = \wp (z)\psi
\ee
Then
\be
\psi ^{\dagger}\psi = {\sigma (x-z+\kappa )\sigma (x+z+\kappa )\over\sigma ^2(x+\kappa )
\sigma (z-\kappa )\sigma (z+\kappa )}
= {\sigma ^2(z)\over\sigma (z +\kappa )\sigma (z - \kappa )}
\left( \wp (z) - \wp (x+\kappa )\right)
\nn \\
\langle \psi ^{\dagger}\psi\rangle =
= {\sigma ^2(z)\over\sigma (z +\kappa )\sigma (z - \kappa )}
\left( \wp (z) - \langle\wp (x+\kappa )\rangle \right)
\ee
and let us take the average over a period $2{\tilde\omega}$ to be
$\langle\wp (x+\kappa )\rangle = 2{\tilde\eta}$. Also
\be\label{qm}
dp = d\left(\zeta (z) + \log\sigma (2{\tilde\omega}-z+\kappa ) - \log\sigma
(\kappa - z)\right) = -dz\left(\wp (z) + \zeta (2{\tilde\omega}-z+\kappa ) -
\zeta (\kappa -z)\right) =
\nn \\
= -dz\left( \wp (z) - 2{\tilde\eta}\right)
\ee
and
\be\label{ppsi}
{dp\over\langle\psi ^{\dagger}\psi\rangle} =
dz{\sigma (z+\kappa )\sigma (z-\kappa)\over\sigma ^2(z)}
\nn \\
d\Omega = {dp\over\langle\psi ^{\dagger}\psi\rangle}\psi ^{\dagger}\psi =
dz{\sigma (x+z+\kappa )\sigma (x+z-\kappa)\over\sigma ^2(z)\sigma ^2(x+\kappa )} =
dz\left(\wp (z) - \wp (x+\kappa )\right)
\ee
Now, the variation $\tilde\delta$ explicitly looks as
\be\label{var1gap}
\tilde\delta d\Omega \equiv {dp\over\langle\psi ^{\dagger}\psi\rangle}
\psi ^{\dagger}_{\kappa}\psi _{\kappa + \delta\kappa} =
dz
{\sigma (z+\kappa )\sigma (z-\kappa)\sigma (x+z+\kappa)
\sigma (x-z+\kappa +\delta\kappa )
\over
\sigma ^2(z)\sigma (x+\kappa)\sigma (z+\kappa)
\sigma (x+\kappa +\delta\kappa )\sigma (z-\kappa -\delta\kappa )} =
\nn \\
= dz{\sigma (x+\kappa +z)\sigma (x+\kappa -z)\over\sigma ^2(z)\sigma ^2(x +\kappa )}
\left[ 1 + \delta\kappa\left(\zeta (x-z+\kappa ) + \zeta (z-\kappa ) -
\zeta (x +\kappa )\right) + {\cal O}\left( (\delta\kappa)^2\right)\right]
\nn \\
= dz\left(\wp (z) - \wp (x +\kappa )\right)
\left[ 1 + \delta\kappa\left(\zeta (x-z+\kappa ) + \zeta (z-\kappa ) -
\zeta (x +\kappa )\right) + {\cal O}\left( (\delta\kappa)^2\right)\right]
\ee
It is easy to see that (\ref{var1gap}) has non-zero residues at $z=0$ and
$z = \kappa $ (the residue at $z = x+\kappa $ is suppressed by
$\wp (z) - \wp (x +\kappa )$. They give
\be\label{res0}
\res _{z=0} \delta d\Omega \sim \delta\kappa \oint _{z\hookrightarrow 0}
dz\wp (z)
\left(\zeta (x-z+\kappa )+\zeta (z-\kappa )\right) \sim
\sim\delta\kappa \oint _{z\hookrightarrow 0}\zeta (z) d\left(\zeta (x-z+\kappa )
+ \right.
\nn \\
\left. +\zeta (z-\kappa )\right)\sim
\delta\kappa\left(\wp (x +\kappa ) +\wp (\kappa )\right)\sim
\delta\left(\zeta (x +\kappa ) + \zeta (\kappa)\right)\equiv \delta v(x)
\ee
and
\be
\res _{z=\kappa }\delta d\Omega = \delta\kappa \left(\wp (\kappa )-\wp (x+\kappa )\right)
= d\Omega (\kappa )
\ee
which follows from the comparison to (\ref{ppsi}).

The quantization of the symplectic structure (\ref{wsform}) is known
to correspond to the complete description of the effective theory
(not only its low-energy part) at least in the simplest case when
$E = W(\mu )$ and $p=Q(\mu )$ were two functions (polynomials) on a
complex sphere. The corresponding generating function (\ref{dS})
was essential in the definition of the duality transformation between two
dual points with completely different behaviour (see \cite{KM} for details).
The exact answer for the partition function $\log{\cal T} = \log{\cal T}_0
+ \log{\cal T}_{\theta} \equiv {\cal F}+ \log{\cal T}_{\theta}$
should also include the deformation of the oscillating
part, corresponding to the {\it massive} excitations.

\section{Whitham equations}

In this section I am going to point out that there exists also another
definition of the generating 1-form (\ref{dS}). This definition goes back
to the general approach to
construction of the effective actions which is known as the
Bogolyubov-\-Whitham
averaging method (see \cite{KriW,Kri,DN} for a comprehensive review
and references).  Though the Whitham dynamics is describes the commutative
flows on the moduli spaces, averaging over the Jacobian -- the fast part
of the theory,
its explicit formulation is most simple and natural in terms of
connections on spectral curves \cite{KriW,Kri}. This definition has an
advantage that it identifies (\ref{prepot}) with the logariphm of
the $\tau $-\-function of the Whitham hierarchy.

\subsection{Whitham equations}

Let us first remind the most general definition of the
Whitham hierarchy given by Krichever \cite{KriW,Kri}.
One has a {\it local} system
of functions $ \Omega _A$ on one-dimensional complex curve and the
corresponding set of parameters $
t_A$ so that it is possible to introduce a $ 1$-form in the space
with co-ordinates $ \left\{ \lambda,t \right\}$ where $\lambda $ is a
{\it local} parameter on a curve
\footnote{Actually it might be better to write
$$
\tilde{\omega} = \omega + \Omega _{\lambda } d\lambda
$$
and consider $ \Omega _{\lambda} = 1$ as a sort of ''gauge condition".}
\begin{equation}\label{1form}
\omega = \sum{\Omega _A \delta t_A}
\end{equation}
The Whitham equations are
\begin{equation}\label{gwhi}
\delta\omega\wedge\delta\omega = 0  \ \ \ \
\delta\omega = \partial _{\lambda} \Omega _A \delta\lambda\wedge\delta t_A +
\partial _B \Omega _A\delta t_A \wedge\delta t_B
\end{equation}
so that one needs to check the independent vanishing of the two
different terms -- $ \delta t ^4$ and $ \delta t^3\delta\lambda$.
The second term gives
\begin{equation}\label{3whi}
\sum{  \partial _\lambda \Omega _A\partial _B \Omega _C} = 0
\end{equation}
for the antisymmetrized sum or introducing explicit co-ordinates
\be\label{t0}
t_{A_0}\equiv x \ \ \ \ \ \ \ \Omega _{A_0}(\lambda ,t)\equiv p(\lambda ,t)
\ee
adjusted to
the fixed choice of parameters $ \left\{ t_A \right\}$
\be\label{pwhi}
\partial _A \Omega _B - \partial _B \Omega _A + \left\{ \Omega _A ,
\Omega _B \right\} = 0
\ \ \
\left\{ \Omega _A , \Omega _B \right\} \equiv {\partial \Omega _A \over
\partial x}{\partial \Omega _B \over \partial p} - {\partial \Omega
_B \over \partial x}{\partial \Omega _A \over \partial p}
\ee
In fact (\ref{pwhi}) strongly depend on the choice of the
local co-ordinate $ p$. The equations (\ref{gwhi}), (\ref{3whi}) and
(\ref{pwhi}) are defined only locally and have huge amount of solutions.

A possible way to restrict ourselves is to get ''globalized picture" related
with the ''modulation"
of parameters of the finite-gap solutions of integrable systems of KP/Toda type.
The KP $\tau$-function
associated with a given spectral curve is
\be\label{KPsol}
{\cal T}\{t_i\} = e^{\sum t_i\gamma _{ij}t_j}\vartheta\left({\bf \Phi}_0
+
\sum t_i{\bf k}_i\right) \ \ \
{\bf k}_i = \oint_{\bf B} d\Omega_i
\ee
where $ \vartheta$ is a Riemann theta-function and $d\Omega_i$ are
meromorphic 1-differentials with poles
of the order $i+1$ at a marked point $z_0$. They are completely
specified by the normalization relations
\be
\oint_{\bf A} d\Omega_i = 0
\label{norA}
\ee
and the asymptotic behaviour
\be
d\Omega_i = \left(\xi^{-i-1} + o(\xi)\right) d\xi
\label{norc}
\ee
where $\xi$ is a local coordinate in the vicinity of $z_0$.
The moduli $\{ u_\alpha \}$ of spectral curve are
invariants of KP flows,
\be
\frac{\partial u_\alpha}{\partial t_i} = 0,
\ee
The way the moduli
depend on $t_i$ after the ''modulation" is defined  by the
Whitham equations (\ref{pwhi}). For $ z = z(\lambda ,t)$ so that
$\partial _i z = \{\Omega _i, z\}$ they acquire more simple form
\be\label{zwhi}
\frac{\partial d\Omega_i(z)}{\partial t_j} =
\frac{\partial d\Omega_j(z)}{\partial t_i}.
\ee
and imply that
\be\label{SS}
d\Omega_i(z) = \frac{\partial dS(z)}{\partial t_i}
\ee
and the equations for moduli, following from (\ref{zwhi}), are:
\be\label{whiv}
\frac{\partial u_\alpha}{\partial t_i} =
v_{ij}^{\alpha\beta}(u)\frac{\partial u_\beta}{\partial t_j}
\ee
with some (in general complicated) functions $v_{ij}^{\alpha\beta}$.

In the KdV (and Toda-chain) case  all the spectral curves are
hyperelliptic, and for the KdV $i$ takes only odd
values $i = 2j+1$, so that
\be
d\Omega_{2j+1}(z) = \frac{{\cal P}_{j+g}(z)}{y(z)}dz,
\ee
the coefficients of the polynomials ${\cal P}_j$ being fixed
by normalization conditions (\ref{norA}), (\ref{norc})
(one usually takes $z_0 = \infty$ and the local parameter
in the vicinity of this point is $\xi = z^{-1/2}$). In this case the
equations (\ref{whiv}) can be diagonalized if the co-ordinates
$ \left\{  u_\alpha \right\}$ on the moduli space are taken to be the
ramification points:
\begin{equation}\label{whihy}
v_{ij}^{\alpha\beta}(u) = \delta ^{\alpha\beta}\left.{d\Omega
_i(z)\over d\Omega _j(z)}\right| _{z=u_\alpha}
\end{equation}
Finally, the differential $dS(z)$ (\ref{SS}) can be constructed for any
finite-\-gap solution \cite{DKN} and it {\it coincides} with the generating
1-form (\ref{dS}). The equality
\be\label{defF}
\frac{\partial {\cal F}}{\partial {\bf a}} = {\bf a}_D \ \ \ \ \ \ \
{\bf a} = \oint_{\bf A} dS \ \ \ \ \ \ \ \ \ {\bf a}_D = \oint_{\bf B} dS
\ee
defines $\tau $-\-function of the Whitham hierarchy
${\cal F} = \log {\cal T}_{Whitham}$ \cite{Kri}. Below, the explicit examples
of the Whitham solutions will be considered.

\subsection{Spherical solutions and topological gravity}

Let us, first, turn to the simplest nontrivial solutions of the systems
(\ref{pwhi}), (\ref{zwhi}) related with $2d$ topological string theories
given at least locally by the formulas
\begin{equation}\label{sphere}
z(\lambda ,t)^p = \lambda ^p + u_{p-2}(t)\lambda ^{p-2} + ... + u_0(t)
\end{equation}
so that
\be\label{spham}
\Omega _i(\lambda ,t) = z(\lambda ,t)^i_+
\ \ \
\Omega _i(z,t) = z^i + O(z^{{i \over n}-1})
\ee
and in the $ p=2$ example
\be\label{p2}
z^2 = \lambda ^2 + U(x,t)
\ \ \
\Omega _i(\lambda ,t) = z(\lambda ,t)^i_{+} =
\left( \lambda ^2 + U(x,t) \right)^{1 \over 2}_+
\nn \\
\Omega _i(z,t) = z^i + O(z^{{i \over 2}-1})
\ee
Now, in the simplest ''global picture" it is possible to interpret
$\lambda $ and $ z$ as global co-ordinates on sphere
with one marked point (where $\lambda = z = \infty$).

For particular choice of parameters $ \left\{ t_k \right\}$
this generic picture gives
Whitham solutions coming from the KP/Toda equations in the following way.
Starting with the ''zero-gap" solution to the KdV hierarchy
\be\label{0gap}
U(x,t) = u = const
\ \ \ \
\Psi (\lambda ,t) = e^{\sum_{k>0}{t_k z^k(\lambda )_+}}
\nn \\
\left( \partial _{t_1}^2 + u \right)\Psi = z^2\Psi
\ \ \ \
\partial _{t_k}\Psi  = \left( \partial _{t_1}^2 + u \right)^{k\over 2}_{+}\Psi
\ee
one comes by
\be\label{average}
\Omega _1 = \overline{\log\Psi (\lambda ,t)_{t_1}} =
z(\lambda )_{+} = \lambda = \sqrt{z^2 - u} \equiv p(z)
\nn \\
\Omega _i = \overline{\log\Psi (\lambda ,t)_{t_i}} = z^{i}(\lambda )_{+} =
\lambda ^3 + {3\over 2}u\lambda = \left( z^2 - u\right)^{3\over 2} +
{3\over 2}u\sqrt{z^2 - u}
\ee
to the formulas (\ref{p2}). The generating differential
(\ref{SS}) is now
\be\label{SKon}
dS(z) = \sum t_k d\Omega _k (z)
\ee
and the solution can be found in terms of the ''periods"
\be\label{resinf}
t_k = {1\over k}\res _{\infty}\left( z^{-k}dS\right)
\ee
For example the third KdV flow gives rise to
\beq\label{whisph}
\partial _{t_3}\Omega _1 (z) = \partial _{t_1}\Omega _3(z)
\eeq
which is equivalent to the Hopf or dispersionless KdV equation
\beq\label{hopf}
u_{t_3} -{3\over 2} uu_{t_1} = 0
\eeq
with generic solution
\be\label{hsph}
t_1 + {3\over 2}t_3u + P(u) = 0
\ \ \ \ \
P(u) \sim \sum t_{2n+1}u^n
\ee
Now, for the purposes of quantum field theory one needs to take from
(\ref{hsph}) the solution of the Whitham equation which is exact solution of
the ''full" KdV
\beq\label{x/t}
u = -{2\over 3}{t_1\over t_3}
\eeq
The generating differential (\ref{SS}) is now
\be\label{dstopgr}
\left. dS \right|_{t_3 = {2\over 3}, t_{k\neq 3} = 0} =
\lambda d\left( z^2 (\lambda )\right) = - (\lambda ^2 + u)d\lambda + d(\dots )
\ee
Then the solution to the linear problem
\beq\label{basph} \left(\partial _{t_1}^2 - {2\over 3}{t_1\over t_3}\right)\Psi
= z^2\Psi
\eeq
gives
\be\label{ai}
\left.\Psi (t_1,z)\right|_{t_3 = {2\over 3}} = Ai \left(
t_1 + z^2 \right)
\nn \\
\phi _i (z) \sim {\partial ^{i-1}\Psi
(z,t)\over\partial t_1^{i-1}} = \sqrt{2z}e^{-{2\over 3}z^3}\int dx
x^{i-1}e^{-{x^3\over 3} + xz^2}
\ee
and the determinant formula
\beq\label{det}
{{\cal T} (t+T)\over{\cal T} (t)} = {\det \phi _i (z_j)\over \Delta (z)}
\eeq
results
in the $\tau $-function of the whole hierarchy in Miwa co-ordinates $T_k = -
{1\over k}\sum z_j^{-k}$.  The decomposition of $\log{\cal T} (T)$ gives
the correlators (\ref{corr}) of two-dimensional topological gravity.

\subsection{Elliptic curves and the Gurevich-Pitaevsky solutions}

Now let us demonstrate that the higher genus Riemann surfaces
(already in the elliptic case) give
rise to nonperturbative formulation of physically less trivial theories.
In contrast to the previous example Whitham times will be nontrivially
related to the moduli of the curve.
For example, from a standard definition (see \cite{KriW,Kri}) in case of the
elliptic curve $y^2 = \prod _{i=1}^3(x - u_i)$ one has
\beq\label{times}
t_k = {2\over k(2-k)}\res (x^{1-{k\over 2}}dy)
\eeq
or substituting $x = \wp (\xi ) + c$ and $y = {1\over
2}\wp '(\xi )$, where $\wp (\xi )$ is the Weierstrass $\wp$-function,
\be\label{times'}
t_k = -{2\over k(2-k)}\res \left( (\wp (\xi ) + c)^{1-{k\over 2}}
\wp ''(\xi )d\xi\right)
= \nn \\
= -{6\over k(2-k)}Res {d\xi\over \xi ^{6-k}}\left( 1+c\xi ^2 + {g_2\over
20}\xi ^4 +
...\right)^{1-{k\over 2}}\left( 1+ {g_2\over 60}\xi ^4 + ...\right) =
\nn \\
= {2\over 5}\delta _{k,5} - c\delta _{k,3} +\left( {3\over 4}c^2 -
{1\over
4}g_2\right)\delta _{k,1} + {\cal O}(t_{-k})
\ee
with dependent negative times $t_{-k} = t_{-k}(t_1,t_3)$ \cite{LGGKM}.

The elliptic solution to the KdV is
\be\label{ukdvsol}
U(t_1,t_3,\ldots| u) =
\frac{\partial^2}{\partial t_1^2} \log{\cal T} (t_1,t_3,\ldots|u) =
\\
= U_0 \wp (k_1t_1 + k_3t_3 + \ldots + \Phi_0 |\omega , \omega ') +
 {u\over 3}
\ee
and
\be
dp \equiv d\Omega_1(z) = \frac{z - \alpha (u)}{y(z)}dz, \nn \\
dQ \equiv d\Omega_3(z) = \frac{z^2 - \frac{1}{2}uz - \beta
(u)}{y(z)}dz.
\label{pp}
\ee
Normalization conditions (\ref{norA}) prescribe that
\be
\alpha (u) = \frac{\oint_A \frac{zdz}{y(z)}}
{\oint_A \frac{dz}{y(z)}}\ \ \ \ \
\beta (u) = \frac{\oint_A \frac{(z^2 - \frac{1}{2}uz)dz}{y(z)}}
{\oint_A \frac{dz}{y(z)}}.
\ee
The simplest elliptic example is the first Gurevich-Pitaevsky (GP) solution
\cite{GP} with the underlying spectral curve
\beq\label{elc1}
y^2 = (z^2 - 1)(z-u)
\eeq
specified by a requirement that all branching points except for $z = u$ are
{\it fixed} and do not obey Whitham deformation.
It is easy to see that by change of variables $z = u + \lambda ^2$
and $y \rightarrow y\lambda$ the curve (\ref{elc1}) can be written as
(a particular $N_c = 2$ case) of the Toda-\-chain curve (\ref{hypelTC})
$y^2 = (\lambda^2 + u)^2 - 1$ c $P_{N_c=2}(\lambda ) = \lambda^2 + h_2$ i.e.
$h_2 \equiv u$.
The generating differential (\ref{dS}), (\ref{SS}),
corresponding to (\ref{elc1}) is given by \cite{GKMMM}
\be\label{dSz}
dS(z) = \left(t_1 + t_3(z+\frac{1}{2}u)  + \ldots \right)
\frac{z-u}{y(z)}dz \ \stackreb{\{ t_{k>1}=0 \} }{=} \
t_1 \frac{z-u}{y(z)}dz
\ee
and it produces the simplest solution to (\ref{zwhi}) coming from the elliptic
curve. From (\ref{dSz}) one derives:
\be
\frac{\partial dS(z)}{\partial t_1} =
\left( z - u - (\frac{1}{2}t_1 + u t_3)
\frac{\partial u}{\partial t_1}\right)\frac{dz}{y(z)}, \nn \\
\frac{\partial dS(z)}{\partial t_3} =
\left( z^2 - \frac{1}{2}uz - \frac{1}{2}u^2 -
(\frac{1}{2}t_1 + u t_3)
\frac{\partial u}{\partial t_3}\right)\frac{dz}{y(z)}, \nn \\
\ldots ,
\ee
and comparison with explicit expressions (\ref{pp}) implies:
\be
(\frac{1}{2}t_1 + u t_3)\frac{\partial u}{\partial t_1} =
\alpha (u) - u, \nn \\
(\frac{1}{2}t_1 + u t_3)\frac{\partial u}{\partial t_3} =
\beta (u) - \frac{1}{2}u^2.
\label{preW}
\ee
In other words, this construction provides the first GP solution to the
Whitham equation
\be\label{GPe}
\frac{\partial u}{\partial t_3} = v_{31}(u)
\frac{\partial u}{\partial t_1},
\ee
with
\be
v_{31}(u) = \frac{\beta(u)-\frac{1}{2}u^2}{\alpha(u)-u} =
\left.\frac{d\Omega_3(z)}{d\Omega_1(z)}\right|_{z=u},
\ee
which can be expressed through elliptic integrals \cite{GP}.

The elliptic solution with all moving branch points is given instead
of (\ref{dSz}) by
\be\label{2gps}
y^2 = \prod _{i=1}^3 (z-u_i)
\ \ \ \ \
dS = y(z)dz
\ee
and all $u_i$ are some functions of the Whitham times. Then one has to
fulfil
\be
dp = {\partial\over\partial t_1}dS = ydz\sum _{i=1}^3{v_i\over z-u_i}=
{z-\alpha (u)\over y}dz
\nn \\
dQ = {\partial\over\partial t_3}dS = ydz\sum _{i=1}^3{w_i\over z-u_i}=
{z^2 - \left( {1\over 2}\sum u_i\right)z -\beta (u)\over y}dz
\ee
where
\be
v_i = -{1\over 2}{\partial u_i\over\partial t_1} \ \ \ \ \ \
w_i = -{1\over 2}{\partial u_i\over\partial t_3}
\ee
and $\alpha (u)$ and $\beta (u)$ are defined as before. Now what one gets
is a simple linear system of the equations
\be
R_{ij}(u)v_j = V_i \ \ \ \
R_{ij}(u)w_j = W_j
\nn \\
R_{ij}(u) = \left(
\begin{array}{ccc}
    1     &     1     &     1     \\
u_2 + u_3 & u_1 + u_3 & u_1 + u_2 \\
  u_2u_3  &   u_1u_3  &   u_1u_2
\end{array}\right)
\nn \\
V^{T} = \left( 0, -1, -\alpha (u)\right)
\ \ \ \
W^{T} = (1, {1\over 2}\left( u_1 + u_2 + u_3), -\beta (u)\right)
\ee
The solution of linear system looks similar to (\ref{GPe}), for example:
\beq
{w_1\over v_1} = {u_1^2 - {1\over 2}(u_1 + u_2 + u_3) - \beta (u)
\over u_1 - \alpha (u)}
\eeq
and this class of solutions contain, for example, pure (non-topological!)
$2d$ gravity. The second GP solution
\footnote{corresponding to the Yang-Lee edge singularity or $(2,5)$ conformal
minimal model interacting with $2d$ gravity}
\be\label{GP2}
y^2 = \prod _{i=1}^3 (z - u_i)
\ \ \ \ \
dS = (z - e)ydz
\ee
can be obtained from elliptic curve with a marked point.
In all these cases, one might expect that a ''modulated" $\tau $-function
(cf. with (\ref{KPsol})) would still have a form
\beq\label{taudef}
{\cal T}\{t_i\} = e^{{\cal F}(t)}\vartheta
\left( {\bf S} (t) \right)\equiv {\cal T}_{Whitham}{\cal T}_{\Theta}
\eeq
so that
\beq
{\bf k}_i(t) = {\partial {\bf S}\over\partial t_i}
\eeq
and the poles of the ''effective" potential
\beq
U(t) = {\partial ^2 \over \partial t_1^2}\log {\cal T}
\eeq
can be identified with the massive excitations (\ref{ymval}).

For the higher genus curves the above procedure looks to be exactly the same
\cite{GKMMM,MartW1,Tak}.
The only problem is in explicit formulas which look nontrivially
already in the GP case, while the formal definitions (\ref{dS}), (\ref{SS})
are again formulated in terms of the eigenvalues of two operators.

\section{Prepotential of the Effective Theory}

In the previous section the prepotential ${\cal F}$ was identified with the
logariphm of the $\tau $-\-function of the Whitham hierarchy. Such
identification, being a particular case of the fundamental formula relating
the generating functions or effective actions of quantum theories to the
$\tau $-\-functions of the hierarchies of integrable equations, has a little
bit implicit form. In this section, I am will formulate and discuss
explicitely a system of differential equations satisfied by the prepotential
${\cal F}$. Following \cite{MMM2} the proof of these equations for the
${\cal N}=2$ SUSY gluodynamics will be presented, the case of ${\cal N}=2$
Yang-\-Mills theory with matter will be discussed in detail in a forthcoming
paper \cite{MMM3}.

\subsection{The associativity equations}

The prepotential ${\cal F}$ \cite{SW1,SW2} is defined in terms of a
family of Riemann surfaces,
endowed with the meromorphic differential
$dS$.  For the gauge group $G=SU(N)$ the family is
\cite{SW1,SW2}, \cite{sun}, \cite{GKMMM} given by (\ref{fsc-Toda})
and the generating differential by (\ref{dS}).
The prepotential ${\cal F}(a_i)$ is implicitly defined by the set of
equations (\ref{defF}).
According to \cite{GKMMM}, this definition identifies
${\cal F}(a_i)$ as logarithm of (truncated) $\tau$-function of
Whitham integrable hierarchy.
Existing experience with Whitham hierarchies (\ref{t0})
\cite{KriW,Dubtop}
implies that ${\cal F}(a_i)$ should satisfy some sort
of the Witten-\-Dijkgraaf-\-Verlinde-\-Verlinde (WDVV) equations
\cite{WDVV}.
Below in this section we demonstrate that WDVV equations
for the prepotential actually look like
\be
{\cal F}_i {\cal F}_k^{-1} {\cal F}_j = {\cal F}_j {\cal F}_k^{-1} {\cal F}_i
\ \ \ \ \ \ \forall i,j,k = 1,\ldots,N-1.
\label{FFF}
\ee
Here ${\cal F}_i$ denotes the matrix
\be
({\cal F}_i)_{mn} = \frac{\partial^3 {\cal F}}{\partial a_i
\partial a_m\partial a_n}.
\ee
Few comments are now in order:\\
{\bf (i)}. Let us remind, first, that the conventional WDVV equations
for topological field theory express the associativity of the algebra
$\phi_i\phi_j = C_{ij}^k\phi_k$ (for symmetric in $i$ and $j$ structure
constants):
$(\phi_i\phi_j)\phi_k = \phi_i(\phi_j\phi_k)$,
or $C_iC_j = C_jC_i$, for the matrix  $(C_i)^m_n \equiv C_{in}^m$.
These conditions become highly non-trivial since, in
topological theory, the structure constants are expressed
in terms of a single prepotential ${\cal F}(t_i)$:
$C_{ij}^l = (\eta^{-1}_{(0)})^{kl}{\cal F}_{ijk}$, and
${\cal F}_{ijk} = {\partial^3{\cal F}\over\partial t_i\partial t_j\partial t_k}$,
while the metric is $\eta_{kl}^{(0)} = {\cal F}_{0kl}$, where $\phi_0 = I$
is the unity operator. In other words, the conventional WDVV equations
can be written as
\be
{\cal F}_i {\cal F}_0^{-1} {\cal F}_j = {\cal F}_j {\cal F}_0^{-1} {\cal F}_i.
\label{FFFconv}
\ee
In contrast to (\ref{FFF}), $k$ is restricted to
$k=0$, associated with the distinguished unity operator.

On the other hand, in the Seiberg-Witten theory there does not clearly exist
any distinguished index $i$: all the arguments $a_i$ of the
prepotential are on equal footing. Thus, if some kind of
the WDVV equations holds in this case, it should be invariant under
any permutation of indices $i,j,k$ -- criterium satisfied by
the system (\ref{FFF}).
Moreover, the same set of equations (\ref{FFF}) is satisfied for
generic topological theory.\\
{\bf (ii)}. In the general theory of Whitham hierarchies
\cite{KriW,Dubtop} the WDVV equations arise also in the form (\ref{FFFconv}).
Again, there exists  a distinguished time-variable (\ref{t0})
for the global solutions usually associated with the first time-variable
of the original
KP/KdV hierarchy. Moreover, usually -- in contrast to the
simplest topological models -- the set of these variables for
the Whitham hierarchy is infinitely large. In this context
our eqs.(\ref{FFF}) state that, for specific subhierarchies
(in the Seiberg-Witten gluodynamics, it is the Toda-chain hierarchy,
associated with a peculiar set of hyperelliptic surfaces),
there exists a non-trivial {\it truncation} of the
quasiclassical $\tau$-function, when it depends on the finite
number ($N-1=g$ -- genus of the Riemann surface) of
{\it equivalent} arguments $a_i$, and satisfies a much wider
set of WDVV-like equations: the whole set (\ref{FFF}).\\
{\bf (iii)}. From (\ref{defF}) it is clear that $a_i$'s
are defined modulo linear transformations
(one can change $A$-cycle for any linear combination
of them). Eqs.(\ref{FFF}) possess adequate ``covariance'':
the least trivial part is that ${\cal F}_k$ can be
substituted by ${\cal F}_k + \sum_l\epsilon_l {\cal F}_l$. Then
\be\label{F-1}
{\cal F}_k^{-1} \rightarrow ({\cal F}_k + \sum \epsilon_l{\cal F}_l)^{-1}
= {\cal F}_k^{-1} - \sum\epsilon_l {\cal F}_k^{-1}{\cal F}_l{\cal F}_k^{-1} +
\sum \epsilon_l\epsilon_{l'} {\cal F}_k^{-1}{\cal F}_l{\cal F}_k^{-1}
{\cal F}_{l'}
{\cal F}_k^{-1} + \ldots
\ee
Clearly, (\ref{FFF}) -- valid for all
triples of indices {\it simultaneously} -- is enough to guarantee that
${\cal F}_i({\cal F}_k + \sum \epsilon_l{\cal F}_l)^{-1}{\cal F}_j =
{\cal F}_j({\cal F}_k + \sum \epsilon_l{\cal F}_l)^{-1}{\cal F}_i$.
Covariance under any replacement of $A$ and $B$-cycles together will be
seen from the general proof below: in fact the role
of ${\cal F}_k$ can be played by ${\cal F}_{d\omega}$, associated with
{\it any} holomorphic 1-differential $d\omega$ on the
Riemann surface.\\
{\bf (iv)}. For metric $\eta$, which is a second derivative,
\be
\eta_{ij} = \frac{\partial^2 h}{\partial a_i\partial a_j}
\equiv h_{,ij}
\ee
(as is the case for our
$\eta^{(k)}_{mn}\equiv ({\cal F}_k)_{mn}$): $\ h = h^{(k)} = \partial
{\cal F}/\partial a_k$),
$\Gamma^{i}_{jk} = \frac{1}{2}\eta^{im}h_{,jkm}$ and the
Riemann tensor
\be
R^i_{jkl} =
\Gamma^i_{jl,k} + \Gamma^i_{kn}\Gamma^n_{jl} - (k \leftrightarrow l) =
\frac{1}{2}\eta^{im}h_{,jklm} -{1\over 4}
\eta^{ip}h_{,pnk}\eta^{nm}h_{,mjl} +
 - (k\leftrightarrow l) =
\nn \\=
- \Gamma^i_{kn}\Gamma^n_{jl} + (k \leftrightarrow l)
= -{1\over 4}
\eta^{im}h_{,mnj}\eta^{np}h_{,pkl}
+ (k\leftrightarrow l)
\ee
In terms of the matrix $\eta = \{(\eta)_{kl}\}$ the
zero-curvature condition $R_{ijkl} = 0$ would be
\be
\eta_{,i}\eta^{-1}\eta_{,j} \stackrel{?}{=}
\eta_{,j}\eta^{-1}\eta_{,i}.
\label{R}
\ee
This equation is remarkably similar to (\ref{FFF}) and
(\ref{FFFconv}), but when $\eta^{(k)}_{ij} = {\cal F}_{ijk}$ is
substituted into (\ref{R}), it contains the {\it fourth}
derivatives of ${\cal F}$:
\be\label{FFFD}
{\cal F}_{k,i}{\cal F}^{-1}_{k}{\cal F}_{k,j}={\cal F}_{k,j}{\cal F}^{-1}_{k}{\cal F}_{i,k}
\ \ \ \ \ \forall i,j,k = 1,\ldots,N-1
\ee
(no summation over $k$ in this formula!), while (\ref{FFF})
is expressed through the third derivatives only.

In ordinary topological theories $\eta^{(0)}$ is always flat,
i.e. (\ref{FFFD}) holds for $k=0$ along with (\ref{FFFconv}) -
and this allows one to choose ``flat coordinates'' where
$\eta^{(0)} = const$. Sometimes -- see the example below --
{\it all} the metrics $\eta^{(k)}$
are flat simultaneously. However, this is not always the case: in the
example of quantum cohomologies of $CP^2$ \cite{KoMa,Manin} eqs.(\ref{FFF})
are true for all $k=0,1,2$, but only $\eta^{(0)}$
is flat (satisfies (\ref{R})), while $\eta^{(1)}$ and
$\eta^{(2)}$ lead to non-vanishing curvatures.\\
{\bf (v)}. It is well known that the conventional WDVV equations (\ref{FFFconv})
are pretty restrictive (at least in the case of so called semisimple
Frobenius manifolds in the terminology of \cite{Dubtop,Manin} which however
do not include one of the most interesting examples of
quantum cohomologies of the Calabi-\-Yau type manifolds with the vanishing
first Chern class): this is an overdetermined system
of equations for a single function ${\cal F}(t_i)$, and it is a kind
of surprise that they possess any solutions at all, and in fact there
exist vast variety of them (associated with Whitham
hierarchies, topological models and quantum cohomologies).
The set (\ref{FFF}) is even more overdetermined than
(\ref{FFFconv}), since $k$ can take {\it any} value. Thus, it is even
more surprising that the solutions still
exist (in order to convince the reader, we supplement the
formal proof by explicit examples).
Of course, (\ref{FFF}) is tautologically true
for $N=2$ and $N=3$, it becomes a non-trivial system for
$N\geq 4$.\\
{\bf (vi)}. Our consideration suggests that when
the {\it ordinary} WDVV (\ref{FFFconv}) is true, the whole system (\ref{FFF})
holds automatically for any other $k$ (with the only restriction that
${\cal F}_k$ is non-\-degenerate).
Indeed,
\footnote{This simple proof was suggested by A.Rosly.}
\be
{\cal F}_i{\cal F}_k^{-1}{\cal F}_j =
{\cal F}_0 \left(C_i^{(0)}(C_k^{(0)})^{-1} C_j^{(0)}\right)
\ee
is obviously symmetric w.r.to the permutation $i\leftrightarrow j$ implied
by $[C_i^{(0)},C_j^{(0)}] = 0$.

This implies that this entire
system should possess some interpretation in the
spirit of hierarchies or hidden symmetries. It still
remains to be found.
The geometrical or cohomological origin of relations (\ref{FFF})
also remains obscure.\\
{\bf (vii)}. One can also look for relation
between (\ref{FFF}) and Picard-Fuchs equations, and then
address to the issue
of the WDVV equations for the prepotential, associated with
families of the Calabi-Yau manifolds.\\
{\bf (viii)}. Effective theory (\ref{defF}) is naively {\it non-\-topological}.
From the 4-dimensional point of view it describes the
low-energy limit of the Yang-Mills theory which -- at least, in
the ${\cal N}=2$ supersymmetric case -- is {\it not} topological and
contains propagating massless particles. Still this theory
is entirely defined by a prepotential, which -- as we now
see -- possesses {\it all} essential properties of the
prepotentials in topological theory. Thus, from the
``stringy'' point of view (when everything is described
in terms of universality classes of effective actions)
the Seiberg-Witten models belong to the same class as
topological models: only the way to extract physically
meaningful correlators from the prepotential is
different. This can serve as a new evidence
that the notion of topological theories is deeper than
it is usually assumed: as emphasized in \cite{GKMMM} it
can be actually more related to the low-energy (IR) limit of
field theories than to the property of the correlation
functions to be constants in physical space-time.

\subsection{The proof of the associativity equations}

Let us start with reminding the proof of the WDVV equations
(\ref{FFFconv}) for ordinary topological theories.
We take the simplest of all possible examples, when
$\phi_i$ are polynomials of a single variable $\lambda$.
The proof is essentially the check of consistency between the
following formulas:
\be
\phi_i(\lambda)\phi_j(\lambda) = C_{ij}^k\phi_k(\lambda)
\ {\rm mod}\ W'(\lambda),
\label{.c}
\ee
\be
{\cal F}_{ijk} = {\rm res}\frac{\phi_i\phi_j\phi_k(\lambda)}
{W'(\lambda)} = \sum_{\alpha}
\frac{\phi_i\phi_j\phi_k(\lambda_\alpha)}
{W''(\lambda_\alpha)},
\label{vc}
\ee
\be
\eta_{kl} = {\rm res}\frac{\phi_k\phi_l(\lambda)}
{W'(\lambda)} = \sum_{\alpha}
\frac{\phi_k\phi_l(\lambda_\alpha)}
{W''(\lambda_\alpha)},
\label{vvc}
\ee
\be
{\cal F}_{ijk} = \eta_{kl}C_{ij}^l.
\label{vvvc}
\ee
Here $\lambda_\alpha$ are the roots of $W'(\lambda)$.

In addition to the consistency of (\ref{.c})-(\ref{vvvc}),
one should know that {\it such} ${\cal F}_{ijk}$, given by
(\ref{vc}), are the third derivatives of a single function ${\cal F}(a)$, i.e.
\be
{\cal F}_{ijk} = \frac{\partial^3{\cal F}}{\partial a_i
\partial a_j\partial a_k}.
\ee
This integrability property of (\ref{vc}) follows from separate
arguments and can be checked independently.
But if (\ref{.c})-(\ref{vvc}) is given, the proof of
(\ref{vvvc}) is straightforward:
\be \eta_{kl}C^l_{ij} = \sum_{\alpha}
\frac{\phi_k\phi_l(\lambda_\alpha)}
{W''(\lambda_\alpha)} C^l_{ij} \stackrel{(\ref{.c})}{=} \\ =
\sum_{\alpha}
\frac{\phi_k(\lambda_\alpha)}
{W''(\lambda_\alpha)} \phi_i(\lambda_\alpha)\phi_j(\lambda_\alpha)
= {\cal F}_{ijk}.
\ee
Note that (\ref{.c}) is defined modulo $W'(\lambda)$,
but $W'(\lambda_\alpha) = 0$ at all the points $\lambda_\alpha$.
Imagine now that we change the definition of the metric:
\be
\eta_{kl} \rightarrow \eta_{kl}(\omega) =
\sum_{\alpha}
\frac{\phi_k\phi_l(\lambda_\alpha)}
{W''(\lambda_\alpha)}\omega(\lambda_\alpha).
\ee
Then the WDVV equations would still be correct, provided the
definition (\ref{.c}) of the algebra is also changed for
\be
\phi_i(\lambda)\phi_j(\lambda) = C_{ij}^k(\omega)\phi_k(\lambda)
\omega(\lambda)\ {\rm mod}\ W'(\lambda).
\label{..c}
\ee
This describes an associative algebra, whenever the
polynomials $\omega(\lambda)$ and $W'(\lambda)$ are co-prime,
i.e. do not have common divisors.
Note that (\ref{vc}) -- and thus the fact that ${\cal F}_{ijk}$
is the third derivative of the same ${\cal F}$ -- remains intact!
One can now take for $\omega(\lambda)$ any of the operators
$\phi_k(\lambda)$, thus reproducing eqs.(\ref{FFF}) for
all topological theories
\footnote{To make (\ref{FFF})
sensible, one should require that $W'(\lambda)$ has only
{\it simple} zeroes, otherwise some of the matrices ${\cal F}_k$
can be degenerate and non-invertible.}.

In the case of the Seiberg-Witten model the polynomials
$\phi_i(\lambda)$ are substituted by the canonical holomorphic
differentials $d\omega_i(\lambda )$ on hyperelliptic surface
(\ref{fsc-Toda}). This surface
can be represented in a standard hyperelliptic form (\ref{hypelTC}).
Instead of (\ref{.c}) and (\ref{..c}) we now put
\be
d\omega_i(\lambda )d\omega_j(\lambda ) =
C_{ij}^k(d\omega) d\omega_k(\lambda )
d\omega(\lambda ) \ {\rm mod}\ \frac{dP_N(\lambda )d\lambda }{y^2}.
\label{.}
\ee
In contrast to (\ref{..c}) we can not now choose $\omega = 1$
(to reproduce (\ref{.c})), because now we need it to be
a 1-differential. Instead we just take $d\omega$ to be a
{\it holomorphic} 1-differential. However, there is no distinguished
one -- just a $g$-parametric family -- and $d\omega$ can be
{\it any} one from this family. We require only that it is
co-prime with $\frac{dP_N(\lambda )}{y}$.

If the algebra (\ref{.}) exists, the structure constants
$C_{ij}^k(d\omega)$ satisfy the associativity condition
(if $d\omega$ and
${dP_N\over y}$ are co-prime). But we still need to show that
it indeed exists, i.e. that if $d\omega$ is given, one can find
($\lambda $-independent) $C_{ij}^k$. This is a simple exercise:
all $d\omega_i$ are linear combinations of
\be
dv_k(\lambda ) = \frac{\lambda ^{k-1}d\lambda }{y}, \ \ \ k=1,\ldots,g: \\
dv_k(\lambda ) = \sigma_{ki}d\omega_i(\lambda ), \ \ \
d\omega_i = (\sigma^{-1})_{ik}dv_k, \ \ \
\sigma_{ki} = \oint_{A_i}dv_k,
\label{sigmadef}
\ee
also $d\omega(\lambda ) = s_kdv_k(\lambda )$.
Thus, (\ref{.}) is in fact a relation between the polynomials:
\be
\left(\sigma^{-1}_{ii'}\lambda ^{i'-1}\right)
\left( \sigma^{-1}_{jj'}\lambda ^{j'-1}\right) =
C_{ij}^k \left(\sigma^{-1}_{kk'}\lambda ^{k'-1}\right)
\left( s_l\lambda ^{l-1}\right) +
p_{ij}(\lambda )P'_N(\lambda ).
\ee
At the l.h.s. we have a polynomial of degree $2(g-1)$.
Since $P'_N(\lambda )$ is a polynomial of degree $N-1=g$, this
implies that $p_{ij}(\lambda )$ should be a polynomial of degree
$2(g-1)-g = g-2$. The identification of two polynomials of
degree $2(g-1)$ impose a set of $2g-1$ equations for the coefficients.
We have a freedom to adjust $C_{ij}^k$ and $p_{ij}(\lambda )$
(with $i,j$ fixed), i.e. $g + (g-1) = 2g-1$ free parameters:
exactly what is necessary. The linear system of equations
is non-degenerate for co-prime $d\omega$ and $dP_N/y$.

Thus, we proved that the algebra (\ref{.}) exists (for a given
$d\omega$) -- and thus $C_{ij}^k(d\omega)$ satisfy the
associativity condition
\be
C_i(d\omega)C_j(d\omega) = C_j(d\omega) C_i(d\omega).
\ee
Hence, instead of (\ref{vc}) we have \cite{KriW,Dubtop,Manin}:
\be
{\cal F}_{ijk} = \frac{\partial^3{\cal F}}{\partial a_i\partial a_j
\partial a_k} = \frac{\partial T_{ij}}{\partial a_k} = \nn \\
= \stackreb{d\lambda =0}{{\rm res}} \frac{d\omega_id\omega_j
d\omega_k}{d\lambda\left(\frac{dw}{w}\right)} =
\stackreb{d\lambda =0}{{\rm res}} \frac{d\omega_id\omega_j
d\omega_k}{d\lambda\frac{dP_N}{y}} =
\sum_{\alpha} \frac{\hat\omega_i(\lambda_\alpha)\hat\omega_j
(\lambda_\alpha)\hat\omega_k(\lambda_\alpha)}{P'_N(\lambda_\alpha)
/\hat y(\lambda_\alpha)}
\label{v}
\ee
The sum at the r.h.s. goes over all the $2g+2$ ramification points
$\lambda_\alpha$ of the hyperelliptic curve (i.e. over the zeroes
of $y^2 = P_N^2(\lambda )-1 = \prod_{\alpha=1}^N(\lambda - \lambda_\alpha)$);
\  $d\omega_i(\lambda) = (\hat\omega_i(\lambda_\alpha) +
O(\lambda-\lambda_\alpha))d\lambda$,\ $\ \ \ \hat y^2(\lambda_\alpha) =
\prod_{\beta\neq\alpha}(\lambda_\alpha - \lambda_\beta)$.
Formula (\ref{v}) can be extracted from \cite{KriW}, and its proof
is presented in \cite{MMM2}.

We define the metric in the following way:
\be
\eta_{kl}(d\omega) =
\stackreb{d\lambda =0}{{\rm res}} \frac{d\omega_kd\omega_l
d\omega}{d\lambda\left(\frac{dw}{w}\right)} =
\stackreb{d\lambda =0}{{\rm res}} \frac{d\omega_kd\omega_l
d\omega_k}{d\lambda\frac{dP_N}{y}} = \\ =
\sum_{\alpha} \frac{\hat\omega_k(\lambda_\alpha)\hat\omega_l
(\lambda_\alpha)\hat\omega(\lambda_\alpha)}{P'_N(\lambda_\alpha)
/\hat y(\lambda_\alpha)}
\label{vv}
\ee
In particular, for $d\omega = d\omega_k$,
$\eta_{ij}(d\omega_k) = {\cal F}_{ijk}$: this choice will
give rise to (\ref{FFF}).

Given (\ref{.}), (\ref{v}) and (\ref{vv}), one can check:
\be
{\cal F}_{ijk} = \eta_{kl}(d\omega)C_{ij}^k(d\omega).
\label{vvv}
\ee
Note that ${\cal F}_{ijk} = {\partial^3{\cal F}\over\partial a_i\partial a_j
\partial a_k}$ at the l.h.s. of (\ref{vvv}) is independent
of $d\omega$! The r.h.s. of (\ref{vvv}) is equal to:
\be
\eta_{kl}(d\omega)C_{ij}^k(d\omega) =
\stackreb{d\lambda =0}{{\rm res}} \frac{d\omega_kd\omega_l
d\omega}{d\lambda\left(\frac{dw}{w}\right)} C_{ij}^l(d\omega)
\stackrel{(\ref{.})}{=} \\ =
\stackreb{d\lambda =0}{{\rm res}} \frac{d\omega_k}
{d\lambda\left(\frac{dw}{w}\right)}
\left(d\omega_id\omega_j - p_{ij}\frac{dP_Nd\lambda}{y^2}\right) =
{\cal F}_{ijk} - \stackreb{d\lambda =0}{{\rm res}} \frac{d\omega_k}
{d\lambda\left(\frac{dP_N}{y}\right)}p_{ij}(\lambda)\frac{dP_Nd\lambda}{y^2}
= \\ = {\cal F}_{ijk} - \stackreb{d\lambda =0}{{\rm res}}
\frac{p_{ij}(\lambda )d\omega_k(\lambda)} {y}
\ee
It remains to prove
that the last item is indeed vanishing for any $i,j,k$.
This follows from the
fact that $\frac{p_{ij}(\lambda )d\omega_k(\lambda )}{y}$
is singular only at zeroes of $y$, it is not singular at
$\lambda =\infty$ because
$p_{ij}(\lambda)$ is a polynomial of low enough degree
$g-2 < g+1$. Thus the sum of its residues at ramification points
is thus the sum over {\it all} the residues and therefore vanishes.

This completes the proof of associativity condition for any $d\omega$.
Taking $d\omega = d\omega_k$ (which is obviously co-\-prime with
$\frac{dP_N}{y}$), we obtain (\ref{FFF}).

\subsection{Explicit examples}

In this section it is demonstrated explicitly that the system (\ref{FFF})
holds for the perturbative part
of the Seiberg-\-Witten prepotential for ${\cal N}=2$ SUSY gluodynamics
and for the quantum cohomologies of $CP^2$. Other examples can be found
in \cite{MMM2}.

Consider first the quantum cohomology of $CP^2$ \cite{KoMa}.
The prepotential is
\be
{\cal F} = \frac{1}{2}t_0t_1^2 + \frac{1}{2}t_0^2t_2 +
\sum_{n=1}^\infty
\frac{N_n t_2^{3n-1}}{(3n-1)!}e^{nt_1}
\ee
and the corresponding matrices are:
\be
{\cal F}_0 = \left(\begin{array}{ccc}
0&0&1\\0&1&0\\1&0&0\end{array}\right) \ \ \
{\cal F}_1 = \left(\begin{array}{ccc}
0&1&0\\1&{\cal F}_{111}&{\cal F}_{112}\\0&{\cal F}_{112}&{\cal F}_{122}
\end{array}\right) \ \ \
{\cal F}_2 = \left(\begin{array}{ccc}
1&0&0\\0&{\cal F}_{112}&{\cal F}_{122}\\0&{\cal F}_{122}&{\cal F}_{222}
\end{array}\right)
\ee
where
\be
{\cal F}_{111} = \sum_n \frac{n^3N_n}{(3n-1)!} t_2^{3n-1}e^{nt_1}\ \ \ \ \ \
{\cal F}_{112} = \sum_n \frac{n^2N_n}{(3n-2)!} t_2^{3n-2}e^{nt_1}\\
{\cal F}_{122} = \sum_n \frac{nN_n}{(3n-3)!} t_2^{3n-3} e^{nt_1} \ \ \ \ \ \
{\cal F}_{222} = \sum_n \frac{N_n}{(3n-4)!} t_2^{3n-4} e^{nt_1}
\ee
One can easily check that every equation in (\ref{FFF})
is true if and only if
\be
{\cal F}_{222} = {\cal F}_{112}^2 - {\cal F}_{111}{\cal F}_{122}.
\label{ququ}
\ee
Indeed,
\be
{\cal F}_1{\cal F}_0^{-1}{\cal F}_2 = \left(\begin{array}{ccc}
0 & {\cal F}_{112} & {\cal F}_{122} \\
{\cal F}_{112} & {\cal F}_{122} + {\cal F}_{111}{\cal F}_{112} & {\cal F}_{222}+{\cal F}_{111}{\cal F}_{122} \\
{\cal F}_{122} & {\cal F}_{112}^2 & {\cal F}_{112}{\cal F}_{122} \end{array}\right), \nn \\
{\cal F}_0{\cal F}_1^{-1}{\cal F}_2 = \frac{1}{{\cal F}_{122}}\left(\begin{array}{ccc}
-{\cal F}_{112} & {\cal F}_{122} & {\cal F}_{222} \\
{\cal F}_{122} & 0 & 0 \\
{\cal F}_{112}^2-{\cal F}_{111}{\cal F}_{122} & 0 & {\cal F}_{122}^2-{\cal F}_{112}{\cal F}_{222}
\end{array}\right), \nn \\
{\cal F}_0{\cal F}_2^{-1}{\cal F}_1 = \frac{1}{{\cal F}_{112}{\cal F}_{222}-{\cal F}_{122}^2}
\left(\begin{array}{ccc}
-{\cal F}_{122} & {\cal F}_{112}^2-{\cal F}_{111}{\cal F}_{122} & 0 \\
{\cal F}_{222} & {\cal F}_{111}{\cal F}_{222}-{\cal F}_{112}{\cal F}_{122} &
{\cal F}_{112}{\cal F}_{222}-{\cal F}_{122}^2 \\
0 & {\cal F}_{112}{\cal F}_{222}-{\cal F}_{122}^2 & 0 \end{array}\right)
\ee
Eq.(\ref{ququ}) is the famous equation,
providing the recursive relations for $N_n$ \cite{KoMa}:
\be \frac{N_n}{(3n-4)!} = \sum_{a+b=n}
\frac{a^2b(3b-1)b(2a-b)}{(3a-1)!(3b-1)!}N_aN_b.
\ee
For example, $N_2=N_1^2$, $N_3 = 12N_1N_2 = 12N_1^3$, $\ldots$

The zero curvature condition (\ref{FFFD}) is obviously
satisfied for $\eta^{0} = {\cal F}_0$: $R_{ijkl}(\eta^{(0)}) = 0$, but
it is not fulfilled already for $\eta^{(1)} = {\cal F}_1$:
\be
R_{1212}(\eta^{(1)}) \sim {\cal F}_{1112}{\cal F}_{1222} - {\cal F}_{1122}^2 =
-N_1^2 e^{3t_1} + \ldots \neq 0.
\ee
Now, let us turn to the leading (perturbative) approximation
to the exact Seiberg-Witten prepotential,  which satisfies (\ref{FFF})
by itself.
The perturbative contribution is non-transcendental, thus
calculation can be performed in explicit form:
\be
{\cal F}_{pert} \equiv {\cal F}(a_i) =
\left.\frac{1}{2}\sum_{\stackrel{m<n}{m,n=1}}^N
(A_m-A_n)^2\log(A_m-A_n)\right|_{\sum_m A_m = 0} = \nn \\
= \frac{1}{2}\sum_{\stackrel{i<j}{i,j=1}}^{N-1}
(a_i-a_j)^2\log(a_i-a_j) +
\frac{1}{2}\sum_{i=1}^{N-1}a_i^2\log a_i
\label{pertF}
\ee
Here we took $a_i = A_i - A_N$ -- one of the many
possible choices of independent variables, which differ by
linear transformations. According to (\ref{F-1})
the system (\ref{FFF}) is covariant under such changes.

We shall use the notation $a_{ij} = a_i - a_j$. The matrix
\be
\{({\cal F}_1)_{mn}\} = \left\{\frac{\partial^3 {\cal F}}{\partial a_1
\partial a_m\partial a_n} \right\} = \nn \\ =
\left(\begin{array}{ccccc}
\frac{1}{a_1} +\sum_{l\neq 1} \frac{1}{a_{1l}} &
-\frac{1}{a_{12}} & -\frac{1}{a_{13}} & -\frac{1}{a_{14}} & \\
-\frac{1}{a_{12}}& +\frac{1}{a_{12}} & 0 & 0 & \\
-\frac{1}{a_{13}}& 0 & +\frac{1}{a_{13}}& 0 &\ldots \\
-\frac{1}{a_{14}}& 0 & 0 &+\frac{1}{a_{14}} & \\
&&\ldots && \end{array}\right)
\ee
i.e.,
\be\label{f}
\{({\cal F}_i)_{mn}\} = \frac{\delta_{mn}(1-\delta_{mi})(1-\delta_{ni})}
{a_{im}} - \frac{\delta_{mi}(1-\delta_{ni})}{a_{in}}
- \frac{\delta_{ni}(1-\delta_{mi})}{a_{im}} + \nn \\ +
\left(\frac{1}{a_i} + \sum_{l\neq i}\frac{1}{a_{ik}}\right)
\delta_{mi}\delta_{ni}
\ee
The inverse matrix
\be\label{f-1}
\{({\cal F}_k^{-1})_{mn}\} = a_k + \delta_{mn}a_{km}(1-\delta_{mk}),
\ee
for example
\be
\{({\cal F}_1^{-1})_{mn}\} = a_1\left(\begin{array}{cccc}
1 & 1 & 1 & . \\ 1 & 1 & 1 & . \\ 1 & 1 & 1 & . \\
&\ldots & & \end{array}\right) +
\left(\begin{array}{cccc}
0 & 0 & 0 & . \\
0 & a_{12} & 0 & . \\
0 & 0 & a_{13} & . \\
& \ldots & & \end{array}\right)
\ee
As the simplest example let us consider the case $N=4$.
We already know that for $N=4$ it is
sufficient to check only one of the eqs.(\ref{FFF}),
all the others follow automatically. We take $k=1$. Then,
\be
{\cal F}_1=\left(
\begin{array}{ccc}
{1\over a_1}+{1\over a_{12}}+{1\over a_{13}}&-{1\over a_{12}}&-{1\over
a_{13}} \\-{1\over a_{12}}&{1\over a_{12}}&0\\-{1\over a_{13}}&0&{1\over
a_{13}}
\end{array}\right)\ \ \ {\cal F}^{-1}_2=\left(
\begin{array}{ccc}
a_2+a_{21}&a_2&a_2\\a_2&a_2&a_2\\a_2&a_2&a_2+a_{23}
\end{array}\right)\\ {\cal F}_3=\left(
\begin{array}{ccc}
{1\over a_{31}}&0&-{1\over a_{31}}\\
0&{1\over a_{32}}&-{1\over a_{32}}\\
-{1\over a_{31}}&-{1\over a_{32}}&{1\over a_3}+{1\over a_{31}}+{1\over
a_{32}}
\end{array}\right)
\ee
and, say,
\be
{\cal F}_1{\cal F}^{-1}_2{\cal F}_3=\left(
\begin{array}{ccc}
\star&-{1\over a_{31}}& \Delta + {a_{21}+a_{23}\over a_{13}^2}\\
-{1\over a_{13}}&\star&{1\over a_{13}}\\
{a_{21}+a_{23}\over a_{13}^2}&{1\over a_{13}}&\star
\end{array}\right)
\ee
where we do not write down manifestly the diagonal terms since, to check
(\ref{FFF}), one only needs to prove the symmetricity of the matrix. This is
really the case, since
\be
\Delta\equiv
{a_2\over a_1a_3}-{a_{21}\over a_1a_{31}}-{a_{23}\over a_3a_{13}} =0
\ee
Only at this stage we use manifestly that $a_{ij}=a_i-a_j$.

Now let us prove (\ref{FFF}) for the general case. We check the equation
for the inverse matrices. Namely, using formulas (\ref{f})-(\ref{f-1}), one
obtains
\be\label{long}
({\cal F}_i^{-1}{\cal F}_j{\cal F}_k^{-1})_{\alpha\beta}=
\\
={a_ia_k\over a_j}+
\delta_{\alpha\beta}(1-\delta_{i\alpha})(1-\delta_{k\alpha})
(1-\delta_{j\alpha}){a_{i\alpha}a_{k\beta}\over a_{j\beta}}+
\delta_{j\alpha}\delta_{j\beta}(1-\delta_{i\alpha})(1-\delta_{k\beta})
\left({1\over a_j}+\sum_{n\ne j}{1\over a_{jn}}\right)+\\
+\delta_{j\alpha}(1-\delta_{i\alpha})a_{i\alpha}\left(
{a_k\over a_j}-{a_{k\beta}\over a_{j\beta}}(1-\delta_{k\beta})
(1-\delta_{j\beta})\right)+\delta_{j\beta}(1-\delta_{k\beta})\left(
{a_i\over a_j}-{a_{i\alpha}\over a_{j\alpha}}(1-\delta_{i\alpha})
(1-\delta_{j\alpha})\right)=\\
={a_ia_k\over a_j}+
\delta_{\alpha\beta}(1-\delta_{i\alpha}-\delta_{k\alpha}
-\delta_{j\alpha}){a_{i\alpha}a_{k\beta}\over a_{j\beta}}+
\delta_{j\alpha}\delta_{j\beta}\left({1\over a_j}+
\sum_{n\ne j}{1\over a_{jn}}\right)+\\
+\delta_{j\alpha}a_{i\alpha}\left(
{a_k\over a_j}-{a_{k\beta}\over a_{j\beta}}(1-\delta_{k\beta}
-\delta_{j\beta})\right)+\delta_{j\beta}\left(
{a_i\over a_j}-{a_{i\alpha}\over a_{j\alpha}}(1-\delta_{i\alpha}
-\delta_{j\alpha})\right)
\ee
where we used that $i\ne j\ne k$. The first three terms are evidently
symmetric with respect to interchanging $\alpha\leftrightarrow\beta$. In
order to prove the symmetricity of the last two terms, we need to use the
identities ${a_k\over a_j}-{a_{k\beta}\over a_{j\beta}}={a_{\beta}a_{jk}\over
a_ja_{j\beta}}\stackrel{k=\beta}{\to}{a_k\over a_j}$,
${a_i\over a_j}-{a_{i\alpha}\over a_{j\alpha}}=
{a_{\alpha}a_{ji}\over a_ja_{j\alpha}}\stackrel{i=\alpha}{\to}{a_i\over
a_j}$. Then, one gets
\be
\hbox{the last line of (\ref{long})}=
\delta_{j\alpha}(1-\delta_{j\beta})
{a_{ij}a_{jk}\over a_j}{a_{\beta}\over a_{j\beta}} +
\delta_{j\beta}(1-\delta_{j\alpha})
{a_{ij}a_{jk}\over a_j}{a_{\alpha}\over a_{j\alpha}} +
\delta_{j\alpha}\delta_{j\beta}{a_ka_{i\alpha}+a_ia_{k\beta}\over a_j}
\ee

It is interesting to note that
in the particular example (\ref{pertF}),
all the metrics $\eta^{(k)}$ are flat. Moreover,
it is easy to find the explicit flat coordinates:
\be
\eta^{(k)} = \eta^{(k)}_{ij}da^ida^j =
{\cal F}_{ijk}da^ida_j = da_ida_j\partial^2_{ij}(\partial_k {\cal F}) = \nn \\ =
\frac{da_k^2}{a_k} + \sum_{l\neq k}\frac{da_{kl}^2}{a_{kl}} = 4\left(
(d\sqrt{a_k})^2 + \sum_{l\neq k}(d\sqrt{a_{kl}})\right).
\ee

\section{Conclusion}

The picture presented above should be actually
considered as a simple version of a generic nonperturbative effective
target-space formulation of {\it string} theory. String theory possesses
a huge amount of "hidden symmetries" which allow one sometimes to determine
the answer without a direct computation. The introduced objects have
a direct generalization for the whole string theory picture where at
the moment only some observations based on consistency requirements for the
relations among dual theories are made \cite{dualn}.
The difference of the presented above picture with generic conception of
string duality is that the above construction is formulated in strict
mathematical sense what still remains to be done for more "rich" string models.

A stringy generalization is straightforward and related first of all with the
prepotentials
arising in the study of realistic models related to the Calabi-Yau
compactifications. All the steps described above can be in principle repeated
leading finally to the integrable models based on the {\it higher-dimensional
complex} manifolds (instead of $1_{\bf C}$-dimensional $\Sigma$). Such
integrable systems are not investigated yet in detail (see however
\cite{Hi,Mark}).

Another problem is that even for the simplest cases considered above the
complete picture still has a lot of open questions. In particular the exact
form of the
full generating function $\log{\cal T}$ is not yet known even for the
Seiberg-\-Witten effective theories. One can also notice that the explicit
examples presented above were usually restricted to the case of $SU(N_c)$
gauge groups and only their simplest representations, in principle there
should exist an invariant language applicable to any gauge group and any
representation.

One more direction is related with the study of effective theories on
(partially) compactified target-\-spaces \cite{SW3,Nekrasov}. The
compactification of one dimension leads to appearance of the well-\-known
class of relativistic integrable models \cite{Ruj} and allows one to
interpret the divisor on a complex curve corresponding to a finite-\-gap
solution in terms of the loop variables. This question, however, deserves
further investigation.

In spite of all the problems it is easy to believe that for all the theories
where it is possible to make any statement about the nonperturbative and
exact quantities there exists something more than a summation of a
perturbation theory. The main idea I tried to advocate above that this
could be the principle of {\it integrability}, which has been checked
already in several examples and based on general belief that the realistic
theory should be a selfconsistent one and adjust automatically its
properties not to be ill-\-defined both at large and small distances.
It looks that an adequate language for the effective formulation of
nonperturbative field and string theories obeying such property can be
looked for among integrable systems.

\section{Acknowledgments}

I am deeply indebted to A.Gorsky, S.Kharchev, I.Krichever, A.Mironov and
A.Morozov for the
fruitful collaboration on various questions discussed above in these notes
and many illuminating discussions. I am grateful to B.Dubrovin,
V.Fock, A.Gerasimov, P.Grinevich, S.Gukov, A.Gurevich,
I.Kolokolov, A.Losev, Yu.Manin, N.Nekrasov, A.Rosly, A.Orlov, V.Rubtsov
and A.Zabrodin for many discussions and explanations and I am also
grateful to M.Bianchi, P.Fre, D.L\"ust, A.Sagnotti, J.Schwarz, I.Tyutin,
B.Voronov and
P.West for interesting discussions concerning relation of the topic of
this paper to more conventional problems of string and field theories.
The work was in
part supported by RFFI grant 96-01-00887 and INTAS grant INTAS-93-2058.
I am grateful to M.Martellini for warm hospitality in Como where this work
was completed.

\end{document}